\documentclass[5p]{elsarticle}

\usepackage{cite}
\usepackage{graphicx}
\usepackage{url}
\usepackage{todonotes}
\usepackage{algorithm,algorithmic}
\usepackage{psfrag,textcomp,tabularx,subfigure}
\usepackage{amsmath}
\usepackage{comment}

\graphicspath{{figure/}{figures/}}

\DeclareMathOperator*{\argmax}{arg\,max}
\DeclareMathOperator*{\argmin}{arg\,min}

\newcommand{\Dc}{\mathcal{D}}
\newcommand{\Lc}{\mathcal{L}}
\newcommand{\Pc}{\mathcal{P}}

\newcommand{\Kc}{\mathcal{K}}
\newcommand{\Mcal}{\mathcal{M}}

\newcommand{\Fig}[1]{Fig.~\ref{fig:#1}}
\newcommand{\Sec}[1]{Sec.~\ref{sec:#1}}
\newcommand{\Tab}[1]{Tab.~\ref{tab:#1}}
\newcommand{\Eq}[1]{(\ref{eq:#1})}

\begin{document}

\begin{frontmatter}

\title{Scheduling of Emergency Tasks for Multiservice UAVs in Post-Disaster Scenarios}

\author[polito]{Cristina~Rottondi}
\author[cnr]{Francesco~Malandrino}
\author[polito]{Andrea~Bianco}
\author[polito,cnr]{Carla~Fabiana~Chiasserini}
\author[athens]{Ioannis~Stavrakakis}

\address[polito]{Politecnico di Torino, Italy}
\address[cnr]{CNR-IEIIT, Italy}
\address[athens]{National and Kapodistrian University of Athens, Greece}

\begin{abstract}

Single-task UAVs are increasingly being employed to carry out surveillance, parcel delivery, communication support, and other specific tasks. 
When the geographical area of operation of single-task missions is common, e.g., in post-disaster recovery scenarios, it is more efficient to have multiple tasks carried out as part of a single UAV mission. In these scenarios, the UAVs' equipment and mission plan must be carefully selected to minimize the carried load and overall resource consumption.
In this paper, we investigate the joint planning of
multitask missions leveraging a fleet of UAVs equipped with a standard
set of accessories enabling heterogeneous tasks. To this end, an
optimization problem is formulated yielding the optimal joint planning
and deriving the resulting quality of the delivered tasks. In
addition, two heuristic solutions are developed for large-scale
environments to cope with the increased complexity of the optimization
framework. The joint planning is applied to a specific scenario of a flood in
the San Francisco area. Results show the effectiveness of the proposed heuristic solutions, which provide good performance and allow for drastic savings in the computational time required to plan the UAVs' trajectories with respect to the optimal approach, thus enabling prompt reaction to the emergency events.

\end{abstract}


\end{frontmatter}

\section{Introduction}
The usage of Unmanned Aerial Vehicles (UAVs) to accomplish different
kinds of tasks in post-disaster recovery scenarios has recently become
the subject of investigation~\citep{erdelj2017help}. Fleets of UAVs performing
environmental monitoring \citep{saeed2017realistic},
dispatching medicines in rural/hardly accessible areas
\citep{bamburry2015drones}, or ensuring mobile connectivity
\citep{fotouhi2018survey} have already been envisioned. As a relevant
example, UAVs are employed in Rwanda to deliver blood packs
to 21 hospitals located in remote and isolated areas on a regular
basis, even in the presence of harsh weather conditions \citep{ackerman2018medical}.

%However, such critical tasks have up to now been considered in isolation, thus requiring separated fleets with equipment, computational resources, and capabilities dimensioned on the specific mission to be performed \citep{lee2017optimization}. 

To date, critical tasks such as those aforementioned exploited dedicated UAVs, with equipment and resources selected 
and dimensioned according to the needs of the specific task 
\citep{lee2017optimization}. However, in a post-disaster scenario, fleets of UAVs are likely 
to be dispatched carrying out different tasks over the same  geographical area. 
For instance, a UAV equipped with a video-camera may be dispatched to monitor a certain geographical area. 
At the same time, another UAV equipped with an antenna may be dispatched to provide communications support over 
a geographical area. If the latter UAV were also equipped with a video-camera  and had 
the needed additional resources (e.g., enough  battery), it could have carried out the surveying task as well, 
eliminating the need for dispatching the former UAV. Thus, a 50\% savings in the number of UAVs would 
be achieved at the cost of some increase in the total required energy, to account for the video-camera 
consumption and the increased load carried by the UAV. Howver, this energy increase would be minimal, 
compared to the energy required to fly the UAV and operate the antenna and would be more than compensated 
for by saving the engagement of another UAV.
Another possible implication of a UAV carrying out two tasks instead of two UAVs carrying out a single task each, 
is that the quality of service associated with each task may be lower in the former case, 
as the UAV is now not dedicated to a specific task only.
For instance, surveillance of some region maybe delayed if providing communication support 
in another region is prioritized. The aforementioned implications of multitasking will be low 
if an effective joint planning of multitask missions is devised. 
In addition, the larger the number of candidate UAVs that can participate in a specific task, 
the lower the impact of multitasking on the quality of the delivered task is expected to be. 

In this study, we investigate a joint
planning of multitask missions leveraging a fleet of UAVs equipped
with a standard set of accessories i) a video monitoring system \citep{kurz2011real}, 
ii) a cellular communication interface, and iii) a mounting frame for parcel carriage, 
to enable UAVs to perform heterogeneous tasks such as medicine/blood delivery, aerial monitoring, 
and mobile connectivity provisioning.
Such planning must take into account the specific constraints generated by every type of mission, 
i.e.: $i)$ limited payload capacity of the UAVs' payload and restricted delivery time windows for the parcel delivery tasks; 
$ii)$ time-varying network connectivity requirements, due to the users' movements within the served area; 
$iii)$ conflicting service needs in the same area (e.g. both monitoring and connectivity provisioning tasks compete for 
the usage of the upload transmission bandwidth provided by an UAV).
Mission plans for each UAV can be crafted at a centralized entity, e.g., the first responders' control center, 
and then transmitted to the individual UAVs via their wireless interface.

The usage of multi-purpose UAVs instead of UAVs specialized for a single mission 
is expected to improve the overall mission service quality, for a given number of employed UAVs.
To show the benefits achieved by the usage of multi-purpose UAVs, we develop an optimization framework based 
on Integer Linear Programming (ILP) to optimally schedule UAVs tasks in a post-disaster environment. 
Due to the necessity of performing heterogeneous tasks with different requirements in terms of data transmission, 
the model attempts to optimize the global task satisfaction level (i.e., the fraction of service requests being 
satisfied by the UAV fleet at every location and time epoch), while imposing that all the delivery tasks are fulfilled within 
their respective time windows.
Since the problem is NP-hard, two heuristic algorithms for larger-scale environments are developed to
cope with the increased complexity of the optimization framework. 
The aim of the two greedy approaches is to ensure fast computation of a feasible solution while 
limiting the required computational complexity, which is a fundamental requirement to enable a fast reaction 
to emergency situations. The case study is a simulated flooding event in the San Francisco area, where UAVs depart 
from one of the depots surrounding the emergency area and must return to a depot after completion of their 
task to change/recharge batteries. 
Results show that our heuristic algorithms provide good performance in comparison to the
optimum, while ensuring rapid calculation of the UAVs' trajectories. Furthermore, fully equipping 
all UAVs, e.g., providing all of them with cameras and radios, implies greater flexibility that outweighs
the resulting lower payload available for parcel delivery missions, further increasing performance. 

The main contributions of the paper can be summarized as follows:
\begin{itemize}
    \item a novel system model, synthetically and effectively describing the main entities involved in multi-task UAV management, along with the decisions to make and their effects;
    \item an optimization formulation of the multi-task UAV management problem, allowing us to assess whether or not multi-task UAVs can attain the same performance of single-purpose ones;
    \item two efficient and effective heuristics, providing different trade-offs between low complexity and high quality decisions.
\end{itemize}

The remainder of the paper is organized as follows. \Sec{related}
briefly reviews the related literature. \Sec{model} presents an
optimization formulation of the multitask UAVs trajectory planning problem. 
We present the heuristic approaches to tackle large instances in \Sec{heuristic}, 
providing a worst-case ratio analysis. We discuss our reference scenario in \Sec{scenario} and 
report our numerical evaluation in \Sec{results}. Conclusions are drawn in the final section.

\section{Related Work}
\label{sec:related}

Beside military and security operations, the usage of UAVs is
envisioned in a plethora of civil applications, ranging from
agriculture to environmental monitoring and disaster management (see
\citep{otto2018optimization} for a thorough taxonomy and survey). According to the categorization of UAV-assisted disaster management frameworks provided in \citep{erdelj2016uav,erdelj2017help}, our work falls under the umbrella of medical applications, as our UAVs can transport and deliver medicines and first-aid items within the served area, and of damage assessment frameworks, as the UAVs are also equipped with monitoring systems to perform video inspections. Moreover, while on flight, the UAVs can be integrated into the emergency communication system if equipped with a cellular communication interface, thus providing connectivity across the disaster area. 

A critical comparison of different types of UAVs to be adopted in emergency scenarios is provided in \citep{erdelj2017help}: the authors conclude that quadcopters offer a good tradeoff between flight autonomy, carried payload weight, and equipment cost. Therefore, in this study, the characteristics of the UAVs we adopt mimic those of typical quadcopters.

In the following, we focus on the three types of tasks encompassed in the scenario under study.

\subsection{UAV placement for wireless coverage}
UAVs can be leveraged in a number of wireless networking applications,
e.g., complementing existing cellular systems by providing additional capacity where needed, or ensuring network coverage in emergency or disaster scenarios (see \citep{mozaffari2018tutorial} for a comprehensive overview). The problem of placing UAVs in terms of height and position in 3D environments to provide wireless coverage has been addressed by several studies: analytical frameworks to compute the optimal height \citep{mozaffari2015UAV} or to maximize the associated revenue \citep{bor2016efficient} have been developed, as well as heuristic approaches \citep{reina2018multi,zhao2018deployment,trotta2018joint}. In \citep{bupe2015relief}, an algorithm controlling the deployment and positioning of UAVs within cells of a mobile network has been implemented and demonstrated using quadcopters. In \citep{hironet}, a two-tier emergency
infrastructure-less network aimed at
providing connectivity in the immediate aftermath of a natural disaster is proposed, where the lower tier connects nearby
survivors in a self-organized mesh via short-range wireless technologies and the upper tier creates long-range wireless links between
UAVs exploring the  area of interest. Several game theoretic approaches for the management of UAV-assisted networks have been proposed \citep{mkiramweni2018game}: the reader is referred to \citep{mkiramweni2019survey} for a comprehensive literature review on game-theoretic techniques adopted in UAV-based wireless networks. Metaheuristic approaches such as local search algorithms \citep{REINA201861} an particle swarm optimization \citep{SANCHEZGARCIA2019129} have been investigated to find UAVs' positions that provide better wireless coverage to the victims in a post-disaster environment.

Differently from the above-mentioned studies, our model
jointly considers a dynamic selection of the areas to be covered depending on the evolution of the disaster over time and based on the predicted mobility patterns of users. Additionally, it
jointly optimizes the scheduling of the UAV moving, covering, monitoring, delivering and recharging actions.

\subsection{UAV-based post-disaster monitoring systems}
As overviewed in \citep{chmaj2015distributed,changchun2010research}, fleets of UAVs operating as distributed processing systems can be adopted as aerial sensing infrastructures for various monitoring tasks including, e.g., surveillance, object detection, movement tracking and support to navigation. In \citep{garapati2017game}, a game-theoretic approach for the distributed scheduling of monitoring and wireless coverage tasks among a fleet of UAVs. In \citep{alfeo2019enhancing}, metaheuristic approaches based on stigmergy and flocking have been proposed to optimize target searches of UAV fleets, in absence of prior knowledge about their location or about positions of potential obstacles.
Prototypes of UAV-based architectures for sensing operations have been described in \citep{zema2017unmanned} and  \citep{luo2015uav} with a specific focus on video capture services. 

In our paper, we consider a conceptually similar UAV equipment of hardware and software modules, where each UAV is equipped with on-board sensors, as well as processing, coordination, and networking capabilities.
Additionally, we tackle some aspects not considered in \citep{zema2017unmanned}, i.e., the scheduling of heterogeneous tasks of multiple UAVs and the integration of the UAVs with a wireless service. With respect to the study in \citep{garapati2017game}, for the task allocation to UAVs, we adopt a centralized scheduling approach instead of a distributed game-theoretic framework.

\subsection{UAVs for parcel delivery}
%Though the usage of UAVs for commercial delivery scopes has not yet become reality,
Several recent studies have already investigated optimization strategies for UAV-assisted delivery models (see \citep{yoo2018drone} for a literature review). In particular, variations of the Traveling Salesman Problem \citep{murray2015flying}, modular optimization approaches \citep{7934790} and heuristic algorithms \citep{peng2019hybrid,krakowczyk2018developing} leveraging UAVs for last-mile delivery have been introduced, also in tandem with traditional truck-based delivery \citep{ferrandez2016optimization,liu2018application}. A game-theoretical framework complemented by machine learning algorithms adopted for local UAV control is proposed in \citep{shiri2019massive} for the path planning under wind perturbation. A comparative study of different metaheuristic approaches for path planning is offered in \citep{ghambari2018comparative}, including particle swarm optimization, artificial bee colony, invasive weed optimization, lightning search and differential evolutionary approaches.

Focusing on medicine delivery, the authors of \citep{scott2017drone}
compare two linear programming models that
combine truck-based transportation and UAV delivery.
In our model, we consider a relatively small geographical area affected by a natural disaster and focus on the last-mile UAV assisted delivery problem, assuming that medicines have been carried beforehand at depots located at the border of the target area.

Finally, note that a preliminary version of this study appears in \citep{malandrino_catania}. In this paper, we provide a more comprehensive discussion on heuristic approaches to tackle the problem, including complexity analysis and proofs of worst-case performance ratio, as well as a performance comparison between two different heuristic algorithms. Such algorithms have been specifically designed to capture the peculiarities of the complex multitasking problem at hand, while maintaining low computational complexity (indeed, they exhibit either linear or quadratic dependencies on the input sizes). We have therefore opted for not leveraging well known metaheuristic approaches such as e.g. genetic algorithms, whose complexity depends not only on the input sizes but also exhibit linear dependency on the number of generations and the population size.

\section{System model and optimization problem}
\label{sec:model}

The notation we use in the remainder of the section is summarized in \Tab{notation}.
Lower-case Greek letters indicate decision variables, lower-case Latin ones indicate parameters. Upper-case, calligraphic Latin letters indicate sets. Upper-case, regular Latin letters with indices indicate a specific element of the corresponding set, e.g., the location of a UAV. Upper-case, regular Latin letters without indices indicate design choices, e.g., UAV range, or system-wide parameters.
All indices are written between parentheses, in lexicographic order.

\paragraph{Space and time}
Time is discretized into a set~$\Kc=\{k\}$ of epochs, while space is discretized in a set~$\Lc=\{l\}$ of locations. The distance between two locations~$l_1$, $l_2$ is indicated as~$v(l_1,l_2)$ (clearly, $v(l,l)=0$). Some locations~$\bar{\Lc}\subseteq\Lc$ host depots.

Importantly, locations can be associated an elevation as well as a latitude and longitude. It follows that, in the scenarios that warrant it, multiple locations can correspond to the same position, at different heights.

Binary variables~$\lambda(d,k,l)$ indicate whether UAV~$d$ is at location~$l$ in epoch~$k$. Clearly, UAVs can only be in one location at a time and, given their maximum speed, in each  epoch they can only travel between locations closer than a maximum distance~$V$. This translates into the following constraints, which hold for any $d\in\Dc$ and $k\in\Kc$:
\begin{equation}
\label{eq:one-location}
\sum_{l\in\Lc}\lambda(d,k,l)=1,
\end{equation}
\begin{equation}
\label{eq:close-travel}
\lambda(d,k,l)\leq\sum_{l'\in\Lc\colon v(l',l)\leq V}\lambda(d,k-1,l')\,\,\,\forall l\in\Lc.
\end{equation}

The combined effect of \Eq{one-location} and \Eq{close-travel} is that drones move from a location to another, without disappearing and without jumping across locations too far away from each other.

\paragraph{Payload}
UAVs have a payload capacity~$Y$ and can carry zero or more payload
items~$p\in\Pc$, each with a mass of~$w(p)$. Examples of payload items (payloads for short) are blood packs or cameras. Binary decision variables $\omega(d,k,p)$ express whether payload~$p$ is carried by UAV~$d$ at time~$k$.
\begin{equation}
\label{eq:capacity}
\sum_{p\in\Pc}w(p)\omega(d,k,p)\leq Y,\quad\forall d\in\Dc,k\in\Kc.
\end{equation}
UAV payload can only change at depot locations; thus, for any $d\in\Dc$, $k\in\Kc$, and $p\in\Pc\colon L(d,k)\notin\hat{\Lc}$, we have:
\begin{equation}
\label{eq:no-payload-change}
\omega(d,k,p)=\omega(d,k-1,p).
\end{equation}

Importantly, \Eq{no-payload-change} implies that, as far as the model is concerned, deliveries are {\em not} dropped at their intended destination; this leads to an overestimation of the UAV weight, hence, its energy consumption. This is intentional, and enables our decisions to account for the fact that some deliveries -- in the worst case, {\em all} deliveries -- may fail, e.g., because local conditions prevent drones from landing. Even in such a worst-case scenario, drones must still have enough energy to go back to the base with all their deliveries onboard.

In \Eq{no-payload-change}, we do not consider the fact
that some payloads, e.g., medicine packs, will be dropped somewhere during the mission. This accounts for the worst-case event that one
or more drops fail, because, e.g., 
ground conditions are not adequate for UAV landing; in such a case, UAVs must have enough energy to bring all payloads back, if need be.

\begin{table}
\caption{Notation
    \label{tab:notation}
} %caption
\footnotesize
\begin{tabularx}{1\columnwidth}{|c|p{1.8cm}|X|}
\hline
Symbol & Type & Meaning \\
\hline\hline
$a(p)\in\Kc$ & parameter & Earliest epoch at which to deliver payload~$p$\\
\hline
$b(p)\in\Kc$ & parameter & Latest epoch at which to deliver payload~$p$\\
\hline
$Y$ & parameter & Payload capacity of UAVs\\
\hline
$D(d,p,k)$ & shorthand & Whether drone~$d$ delivers payload~$p$ at time~$k$\\
\hline
$E$ & parameter & Battery capacity of UAVs\\
\hline
$e_v$ & parameter &
 Energy consumed for the vertical ascent and descent when making a delivery
 \\
\hline
$e(l_1,l_2)$ & parameter & Energy consumed when traveling between locations~$l_1$ and~$l_2$, per unit of weight\\
\hline
$f(p)\in\Lc$ & parameter & Location at which payload~$p$ shall be delivered\\
\hline
$\Kc$ & set & Epochs\\
\hline
$\Lc$ & set & Locations\\
\hline
$\hat{\Lc}\subseteq\Lc$ & set & Locations with depots\\
\hline
$L(d,k)\in\Lc$ & shorthand & Location of UAV~$d$ at epoch~$k$\\
\hline
$\Mcal$ & set & Non-delivery missions, e.g., coverage or monitoring\\
\hline
$n(k,m,l)$ & parameter & Work per epoch for mission~$m$ needed by users in location~$l$ \\
\hline
$q(l,m)$ & parameter & Work quality for mission~$m$ that a UAV at location~$l$\\
\hline
$r(m,p)\in\{0,1\}$ & parameter & Whether payload~$p$ is necessary to perform mission~$m$\\
\hline
$s(m)$ & parameter & Data generated by performing one unit of work of mission~$m$\\
\hline
$\Pc$ & set & Payload items\\
\hline
$\hat{\Pc}\subseteq\Pc$ & set & Payload items to be delivered\\
\hline
$t(l_1,l_2)$ & parameter & Whether traffic can be transferred between UAVs at~$l_1$ and~$l_2$\\
\hline
$T$ & parameter & Capacity (land-to-air) of the UAVs' on-board base station\\
\hline
$V$ & parameter & Maximum distance a UAV can cover in one epoch\\
\hline
$W$ & parameter & UAV weight\\
\hline
$w(p)$ & parameter & Weight of payload~$p$\\
\hline
$v(l_1,l_2)$ & parameter & Distance between locations~$l_1$ and~$l_2$\\
\hline
$\beta(d,k)$ & real variable & Battery level of UAV~$d$ at epoch~$k$\\
\hline
$\lambda(d,k,l)$ & binary variable & Whether UAV~$d$ is in location~$l$ at epoch~$k$\\
\hline
$\mu(d,k,m)\in[0,1]$ & real variable & Fraction of epoch~$k$ that UAV~$d$ spends in mission~$m$\\
\hline
$\sigma(k,m,l)\in[0,1]$ & real aux. variable & Satisfaction of users in location~$l$ concerning mission~$m$ at epoch~$k$\\
\hline
$\tau(d,k)$
& binary variable & Whether traffic generated at UAV~$d$ at epoch~$k$ can reach the cellular network\\
\hline
$\omega(d,k,p)$ & binary variable & Whether UAV~$d$ carries payload~$p$ at epoch~$k$\\
\hline
\end{tabularx}
\end{table}

\paragraph{Energy and battery}
Real variables~$\beta(d,k)$ express the battery level of UAV~$d$ at epoch~$k$. Clearly, such variables must be positive and can never exceed the battery capacity~$E$, i.e.,
\begin{equation}
\label{eq:b-range}
0\leq \beta(d,k)\leq E,\quad\forall d\in\Dc,k\in\Kc.
\end{equation}
Next, we need to account for power consumption.
Indicating with~$D(d,k,p)\in\{0,1\}$ whether drone~$d$ delivers payload~$p$ at time~$k$, we can write:
\begin{multline}
\label{eq:b-consumption}
\beta(d,k)\leq \beta(d,k-1) -e_v\sum_{p\in\hat{\Pc}} D(d,p,k)\\
-e(L(d,k-1),L(d,k))\left(W+\sum_{p\in\Pc}\omega(d,k,p)w(p)\right),\\
\quad\forall d\in\Dc,k\in\Kc\colon L(d,k)\notin\hat{L}.
\end{multline}
In \Eq{b-consumption}, the energy consumed at time~$k$ is given by the product between a factor~$e(l_1,l_2)$, 
accounting for the distance between the locations, i.e., for how far the UAV had to travel, the total weight of the UAV,
plus the vertical energy consumption~$e_v$ incurred if the drone performs a delivery at time~$k$.
Importantly, $e(l,l)>0$, i.e., energy is also consumed by hovering over the same location. A UAV  weight is given by the weight~$W$ of the UAV itself and the sum of the weight of the payload items it carries. Note that \Eq{b-consumption} does not hold at depot locations in~$\hat{\Lc}$, as there UAVs can recharge or swap their batteries.
The actual values of the~$e(l_1,l_2)$ parameters can be set according to any energy model, e.g.,~\citep{zeng2019energy}.
Such energy models are also able to account for the energy consumption due to elevation changes across locations, if any, and to embed such a consumption in the values of~$e(l_1,l_2)$.
Also, notice that the energy consumed to transmit/receive data is not accounted for in our model, as it is negligible to the energy needed to move the UAV.

\paragraph{Delivery missions}
Some payload items~$\hat{\Pc}\subseteq\Pc$ must be delivered at certain locations and times. Specifically, parameters~$f(p)\in\Lc$, $a(p)\in\Kc$, $b(p)\in\Kc$ indicate the target location (final point), as well as the earliest and latest times at which the delivery can take place: the time interval defined by the earliest and latest times will define the delivery time window for a payload. The following constraint imposes that all deliveries are carried out:
\begin{equation}
\label{eq:delivery}
\sum_{d\in\Dc}\sum_{k=a(p)}^{b(p)}\omega(d,k,p)\lambda(d,k,f(p))\geq 1,\quad\forall p\in\hat{\Pc}.
\end{equation}
Eq.\,\Eq{delivery} can be read as follows: there must be at least one epoch between~$a(p)$ and~$b(p)$ during which a UAV~$d$ visits the target location~$f(p)$ while carrying payload~$p$.
It is important to stress that associating a constraint with delivery missions also implies that such missions {\em must} be performed, and any additional tasks will be undertaken only if it does not interfere with deliveries.

\paragraph{Additional missions}
We consider a set~$\Mcal=\{m\}$ of additional missions, e.g., wireless network coverage and video monitoring. We express the demand of each location~$l$ for mission~$m$ at epoch~$k$ through parameters~$n(k,m,l)$, e.g., the traffic offered by the users\footnote{For simplicity and without loss of generality, in this paper we focus on uplink traffic.}. Parameters~$q(l,m)$ express how well a UAV in location~$l$ can perform mission~$m$, e.g., the quality of coverage it can provide. Furthermore, parameters~$r(m,p)\in\{0,1\}$ express the fact that some payload items~$p$, e.g., radios, are needed for mission~$m$. Finally, parameters~$s(m)$ express how much data is generated by performing one unit of work in mission~$m$.
The values of the~$n(k,m,l)$ parameters can be, depending upon the situation, estimated {\em a priori} (e.g., the escape roads from a given area are routinely planned in advance) and/or detected during the disaster (e.g., the areas closer to the coast need to be monitored during/after a flooding).

The main decision to make is how long UAVs perform additional missions. This is conveyed by variables $\mu(d,k,m)\in[0,1]$, expressing the fraction of epoch~$k$ that UAV~$d$ uses to perform mission~$m$. The first constraint we  impose is that UAVs do not perform missions that they are not equipped for:
\begin{multline}
\label{eq:mission-equipment}
\mu(d,k,m)\leq \omega(d,k,p)\\
\quad\forall d\in\Dc,k\in\Kc,m\in\Mcal,p\in\Pc\colon r(m,p)=1.
\end{multline}
Also, we cannot exceed the need of locations:
\begin{multline}
\label{eq:mission-needs}
\sum_{d\in\Dc}\mu(d,k,m)q(L(d),m)\leq n(k,m,l)\\
\quad\forall k\in\Kc,m\in\Mcal, l\in\Lc.
\end{multline}
Note that \Eq{mission-needs} also accounts for the quality with which UAVs at different locations can perform the missions, e.g., the maximum network capacity or monitoring capability that a UAV in location~$l$ can offer.

Next, we need to ensure that all the data traffic generated by
additional missions is transferred  to the in-field deployed cellular
network (denoted with $\Omega$), so that it can be offloaded to the backbone network
infrastructure.
To this end, we denote with~$t(l_1,l_2)\in\{0,1\}$ whether a UAV in location~$l_1$ and one in location~$l_2$ can exchange data, and by~$t(l,\Omega)\in\{0,1\}$ whether a UAV in location~$l$ can send data to the in-field cellular network.
Such information can be obtained from existing models and/or experimental measurements, as better detailed in \Sec{scenario} later. A value of~$t(l_1,l_2)=0$ corresponds to the fact that it is impossible to send any data between~$l_1$ and~$l_2$, e.g., because the locations are too far away.

Given the above, we can determine whether the data generated by additional missions performed by UAV~$d$ at time~$k$ can make its way to the cellular network, through a binary variable~$\tau(d,k)$:
\begin{equation}
\label{eq:tau}
\tau(d,k)\leq t(L(d,k),\Omega)+\sum_{d'\neq d} \tau(d',k)t\left(L(d,k),L(d',k)\right).
\end{equation}
Eq.\,\Eq{tau} says that UAV~$d$ can route its data to the cellular network if~$\Omega$ can be reached directly from its location (first term), or if~$d$ can reach another UAV~$d'$ that, in turn, can reach the cellular network (i.e.,~$\tau(d',k)=1$).
Given the~$\tau$ values, we impose that no UAV performs any mission that generates data, unless such data can reach the network:
\begin{equation}
\label{eq:nodatanomission}
\sum_{m\in\Mcal}\mu(d,m,k)s(m)\leq\tau(d,k)T,\quad\forall d\in\Dc,k\in\Kc.
\end{equation}
Eq.\,\Eq{nodatanomission} also imposes that the quantity of data generated by all missions performed by UAV~$d$ at epoch~$k$ does not exceed the quantity~$T$ of data that can be transferred through the onboard radio interface.

\paragraph{Objective and complexity}
As a first step, we define the satisfaction~$\sigma(k,m,l)$ of location~$l$ at epoch~$k$ for mission~$m$. Such a value is the ratio between the amount of service that the location was actually provided over that needed:
\begin{equation}
\label{eq:sigma}
\sigma(k,m,l)=\frac{\sum_{d\in\Dc}\mu(d,k,m)q(L(d),m)}{n(k,m,l)}.
\end{equation}
We then define our objective as maximizing the sum of the $\sigma$-values defined in \Eq{sigma}:
\begin{equation}
\label{eq:obj}
\Theta = \max \sum_{k\in\Kc,m\in\Mcal,l\in\Lc}\sigma(k,m,l).
\end{equation}

The problem of optimizing \Eq{obj} subject to the constraints \Eq{one-location}--\Eq{sigma} is combinatorial in nature and, like many combinatorial problems, is NP-hard. Indeed, deciding the payload of each UAV in each trip can be seen as an instance of bin-packing -- which is itself an NP-hard problem~\citep{man1996approximation} --, subject to additional constraints concerning energy and delivery times. It follows that it is impractical to directly solve the problem for all but small instances; therefore, we devise several heuristic algorithms as detailed next.

\section{Heuristic Algorithms}
\label{sec:heuristic}

In this section we present two heuristic algorithms aimed at tackling large instances of the considered post-disaster scenario: the first one is a low-complexity greedy algorithm, whereas the second one has higher complexity and builds upon the insertion method first proposed in \citep{solomon1987algorithms}.

Both heuristic approaches leverage a graph-based representation of the considered scenario, where every delivery location $f(p) \in \Lc \colon p \in \hat{\Pc}$ is identified by a graph node $l(p)$. Additional nodes are added to identify the depots in $\hat{\Lc}$. Here, for simplicity, we consider the case of a single depot, i.e.,$\hat{\Lc}=\{l^*\}$. Arc $(l,l')_{g}$ represents route $g\in G_{ll'}$ connecting delivery locations $l(p)$ and $l'(p')$ 
(to simplify the notation, the identifier $(p)$ is omitted). Each route $g$ consists of a list $\nu_g^{ll'}=[l_1,...,l_{\psi(l,l')_g}]$ of locations in $\mathcal{L}$ that are visited at each time epoch while traveling from $l(p)$ to $l'(p')$, where $\psi(l,l')_g$ is the travel duration, in number of epochs (clearly, $l_1=l(p)$ and $l_{\psi(l,l')_g}=l'(p')$). Note that hovering over the same location $l_i$ is captured in the list by replicating $l_i$ once per every hovering epoch. In the remainder of the section, we will indicate the location visited during the $k$-th epoch of route $g\in G_{ll'}$ as $\nu^{ll'}_g(k)$, with $k=1,...,\psi(l,l')_g$.
Note that we consider a set $G_{ll'}$ of alternative routes
for each location pair, which include the $\xi$ shortest paths\footnote{Note that opting for longer lightpaths could increase the satisfaction level of additional missions, but might as well lead to infeasibilities of the delivery missions due to violation of energy or delivery windows constraints.} (i.e., $|G_{ll'}|=\xi$, where $\xi$ is a system parameter).

\begin{figure}
\centering
\includegraphics[width=1\columnwidth]{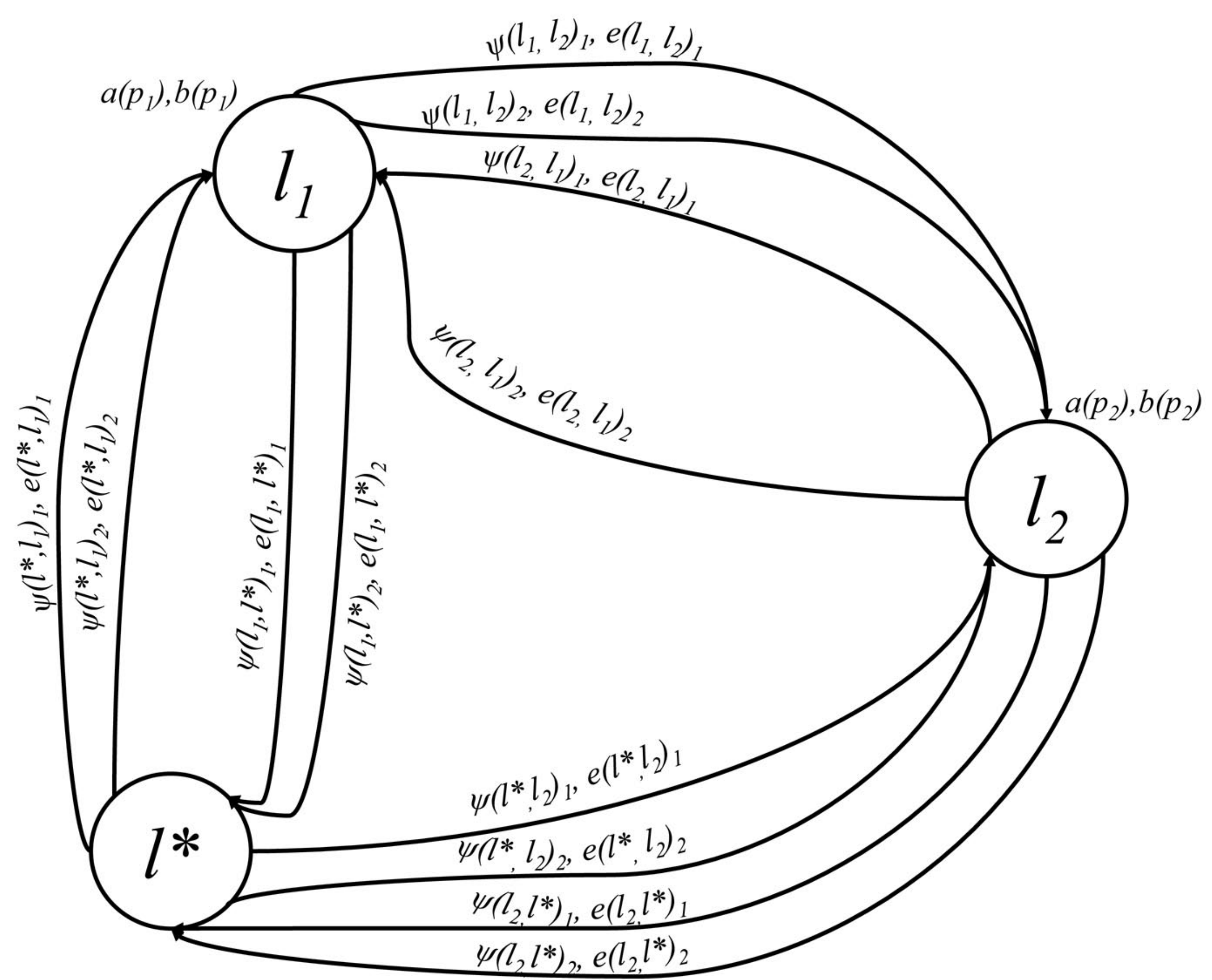}
\caption{Example of graph topology with two nodes $l_1,l_2$ and a depot node $l*$. Arc weights and delivery time windows are also highlighted.}\label{fig:graph} 
\end{figure}

Each arc is associated with two weights: the travel duration $\psi(l,l')_g$ and the energy $e(l,l')_g$ spent by the UAV to travel from node $l$ to $l'$ along route $g \in G_{ll'}$. Energy is computed under the conservative assumption that the UAV's payload is always fully loaded. An example of graph topology with one depot and two delivery nodes is reported in Figure \ref{fig:graph}.

Note that both algorithms determine a single tour for each UAV. Consecutive tours can be obtained by running the algorithms for $J$ consecutive non-overlapping time horizons $\Kc_1,...,\Kc_J$, where time horizons are sized considering the maximum number of epochs a UAV can fly, either hovering or cruising among locations, without being recharged. In this case, the set of deliveries $\hat{\Pc}$ is split into subsets $\hat{\Pc_1},...\hat{\Pc_J}$, such that $\cup_{j=1}^{J}\hat{\Pc_j}=\hat{\Pc}$, where each set $\hat{\Pc_j}$ contains the deliveries $ p \in \hat{\Pc} \colon \sum_{i=1}^{j-1}|\Kc_i| \leq b(p) \leq \sum_{i=1}^{j}|\Kc_i|$.

\subsection{Greedy heuristic}
\begin{algorithm}[tb]
 \caption{Greedy Algorithm}
    \label{alg:greedy}
\begin{algorithmic}[1]
\begin{small}
\STATE on input of $\mathcal{D}$; $\mathcal{K}$; $\mathcal{M}$; $\mathcal{L}$; $\hat{\mathcal{P}}$; $E$; $T$; $w(p),a(p),b(p), f(p) \forall p \in \hat{\mathcal{P}}$; $l^*$; $\psi(l,l')_g,e(l,l')_g,\nu^{ll'}_g, \forall l=f(p),l'=f'(p') \in \Lc, g \in G_{ll'}$; $\Delta(d),$ $\alpha_{md},$ $ n(k,m,l),$ $ q(l,m)$, $s(m)$ $ \forall m \in \mathcal{M},$ $ d \in \mathcal{D},$ $ l \in \mathcal{L}$; 
\STATE create ordered list $X$ of items to be delivered in ascending order of $b(p)$
\FORALL {$d \in \mathcal{D}$}
	\STATE $TourL_d \leftarrow [l^*]$, $TourA_d \leftarrow []$, $e_d \leftarrow 0, \pi_d \leftarrow 0, h_d \leftarrow 0, \overline{l} \leftarrow l^*$
	\FORALL {$p \in X $}
	\STATE $\hat{l} \leftarrow l(p);$ find best route $\overline{g}$ from location $\overline{l}$ to location $\hat{l}$
	\IF{inserting location $\hat{l}$ in $TourA_d$ is feasible}
	\STATE Append arc $(\overline{l},\hat{l})_{\overline{g}}$ to $TourA_d$ and location $\hat{l}$ to $TourL_d$,  $e_d \leftarrow e_d+e(\overline{l},\hat{l})_{\overline{g}}, \pi_d \leftarrow \pi_d+\psi(\overline{l},\hat{l})_{\overline{g}}, h_d \leftarrow h_d+w(p), \overline{l} \leftarrow \hat{l},$  remove $p$ from $X$
%	\FORALL{$k=1,...,\psi(\overline{l},\hat{l})_{\overline{g}},l \in \nu_g^{\overline{l}\hat{l}},m \in \mathcal{M}$}
%	\STATE update parameters $n(\pi_d+k,m,l)$ 
%	\ENDFOR
	\STATE $\forall m \in \Mcal, k \in \Kc, l \in \Lc \ \mu(d,m,k),n(k,m,l) \leftarrow$ \texttt{update}($\mathcal{K},\mathcal{M},\mathcal{L},T,\pi_d$ ,$\psi(\overline{l},\hat{l})_{\overline{g}},\nu_{\overline{g}}^{\overline{l}\hat{l}},$ $\alpha_{md},$ $ n(k,m,l),$ $ q(l,m),$ $s(m)$ $ \forall m \in \mathcal{M},$ $ l \in \mathcal{L}, k \in \Kc$)
	\ENDIF
	\ENDFOR
	\IF{$TourL_d$ contains at least one location different than $l^*$}
		\STATE find best route $\overline{g}$ from location $\overline{l}$ to location $l^*$ 
		\STATE Append arc $(\overline{l},l^*)_{\overline{g}}$ to $TourA_d$ and location $l^*$ to $TourL_d$
%	\FORALL{$k=1,...,\psi(\overline{l},l^*)_{\overline{g}} ,m \in \mathcal{M}$}
%	\STATE update parameters $n(\pi_d+k,m,\nu_g^{\overline{l}l^*}(k))$ 
%	\ENDFOR
\STATE $\forall m \in \Mcal, k \in \Kc, l \in \Lc \ \mu(d,m,k),n(k,m,l) \leftarrow $\texttt{update}($\mathcal{K},\mathcal{M},\mathcal{L},T,\pi_d$ ,$\psi(\overline{l},l^*)_{\overline{g}},\nu^{\overline{l}l^*}_{\overline{g}},$ $\alpha_{md},$ $ n(k,m,l),$ $ q(l,m),$ $s(m)$ $ \forall m \in \mathcal{M},$ $ l \in \mathcal{L}, k \in \Kc$)
	\ENDIF
\ENDFOR
\IF{$X \neq \oslash$ }
 \RETURN no feasible solution found
\ELSE
\STATE $\forall d \in \Dc, m \in \Mcal, k \in \Kc \ \mu(d,m,k) \leftarrow$ \texttt{conn\_ check}($\mathcal{K}$; $\mathcal{D}$; $\mathcal{M}$; $\mathcal{L}$; $TourA_d \forall d \in \Dc$; $t(l,l') \forall l,l' \in \mathcal{L}$; $\mu(d,m,k) \forall d \in \Dc, m \in \Mcal, k \in \Kc$)
\RETURN {$\mu(d,m,k),TourL_d,TourA_d  \forall d \in \mathcal{D}, m \in \Mcal, k \in \Kc$}
\ENDIF
\end{small}
\end{algorithmic}
\end{algorithm}

\begin{algorithm}[tb]
 \caption{\texttt{update} routine (updates mission requirements per epoch)}
    \label{alg:update}
\begin{algorithmic}[1]
\begin{small}
\STATE on input of $\mathcal{K}$; $\mathcal{M}$; $\mathcal{L}$; $\psi(l,l')_g,\nu^{ll'}_g,$ $\alpha_{md},$ $ n(k,m,l),$ $ q(l,m),$ $s(m)$ $ \forall m \in \mathcal{M},$ $ l \in \mathcal{L}, k \in \Kc$, $T$, $\pi_d$; 
\FORALL {$k =1,..., \psi(l,l')_g$}
\IF {$\sum_{m \in \Mcal} q(\nu^{ll'}_g(k),m)s(m)\leq T$}
\FORALL {$m \in \Mcal$}
\STATE $\mu(d,m,\pi_d+k) \leftarrow 1$, $n(\pi_d+k,m,\nu_g^{ll'}(k)) \leftarrow \max(0,n(\pi_d+k,m,\nu_g^{ll'}(k))-q(\nu^{ll'}_g(k),m))$
\ENDFOR
\ELSE
\FORALL {$m \in \Mcal \colon \alpha_{md}>0$}
\STATE $\mu(d,m,\pi_d+k) \leftarrow \frac{\alpha_{md}s(m)}{\sum_{m' \in \Mcal} \alpha_{m'd}s(m')}$, $n(\pi_d+k,m,\nu_g^{ll'}(k)) \leftarrow \max(0,\frac{\alpha_{md}s(m)}{\sum_{m' \in \Mcal} \alpha_{m'd}s(m')}n(\pi_d+k,m,\nu_g^{ll'}(k))-q(\nu^{ll'}_g(k),m))$
\ENDFOR
%\FORALL {$m \in \Mcal \colon \alpha_{md}=0$}
%\STATE $\mu(d,m,\pi_d+k) \leftarrow \frac{s(m)}{\sum_{m' \in \Mcal} s(m')}$, $n(\pi_d+k,m,\nu_g^{ll'}(k)) \leftarrow \max(0,\frac{s(m)}{\sum_{m' \in \Mcal} s(m')}n(\pi_d+k,m,\nu_g^{ll'}(k))-q(\nu^{ll'}_g(k),m))$
%\ENDFOR
\ENDIF
\ENDFOR
\RETURN  $\mu(d,m,k), n(k,m,l) \forall d \in \Dc, m \in \Mcal, k \in \Kc l \in \Lc$
\end{small}
\end{algorithmic}
\end{algorithm}

\begin{algorithm}[tb]
 \caption{\texttt{conn\_ check} routine (checks connectivity among UAVs)}
    \label{alg:check}
\begin{algorithmic}[1]
\begin{small}
\STATE on input of $\mathcal{K}$; $\mathcal{D}$; $\mathcal{M}$; $\mathcal{L}$; $TourA_d \forall d \in \Dc$; $t(l,l') \forall l,l' \in \mathcal{L}$; $\mu(d,m,k) \forall d \in \Dc, m \in \Mcal, k \in \Kc$; 
\FORALL {$d  \in \Dc$, $k \in \Kc$}
\STATE $c_d \leftarrow 0$
\FORALL {$d' \in \Dc \colon d' \neq d$}
\IF{$t(L(d,k),\Omega)>0 \vee \left( t(L(d,k),L(D',k'))>0\wedge t(L(d',k),\Omega)>0\right)$}
\STATE $c_d \leftarrow 1$
\ENDIF
\ENDFOR
\IF{$c_d=0$}
\STATE $\mu(d,m,k)\leftarrow 0 \forall m \in \Mcal$
\ENDIF
\ENDFOR
\RETURN  $\mu(d,m,k) \forall d \in \Dc, m \in \Mcal, k \in \Kc$

\end{small}
\end{algorithmic}
\end{algorithm}

\begin{table*}
\begin{equation}\label{eq:g_ins}
\overline{g}=  \min_{g \in G_{l_1l_2}} (1-\sum_{m \in \mathcal{M}}\alpha_{md})\psi(l_1,l_2)_g - \left(\sum_{m \in \mathcal{M}}\alpha_{md} \sum_{k=1,...,\psi(l_1,l_2)_g}n(\pi_d+k,m,\nu^{l_1l_2}_g(k))q(\nu^{l_1l_2}_g(k),m)\right)
\end{equation}
\begin{align}\label{eq:phi1_ins}
\phi_1(l_{i-1},\overline{l},l_{i})= & \min_{g \in G_{l_{i-1},\overline{l}},g' \in G_{\overline{l},l_{i}} } \left(1-\sum_{m \in \mathcal{M}}\alpha_{md}\right) \cdot \left(\psi(l_{i-1},\overline{l})_{g}+\psi(\overline{l},l_{i})_{g'}-\psi(l_{i-1},l_{i})_{\hat{g}}\right)+\\ \nonumber
& -\sum_{m \in \mathcal{M}}\alpha_{md}\left( \sum_{k=1}^{\psi(l_{i-1},\overline{l})_g}n(\pi_d+k,m,\nu^{l_{i-1}\overline{l}}_g(k))\cdot q(\nu^{l_{i-1}\overline{l}}_g(k),m)+\sum_{k=1}^{\psi(\overline{l},l_{i})_{g'}}n(\pi_d+k,m,\nu^{\overline{l},l_{i}}_{g'})\cdot q(\nu^{\overline{l},l_{i}}_{g'}(k),m)+\right.\\ \nonumber
& +\sum_{j=i+1}^u\sum_{k=1}^{\psi(l_{j-1},j)_{\hat{g}}}n(\pi_d+\psi(\overline{l},l_i)_g'+\sum_{y=i+1}^{j-1}\psi(l_{y-1},l_y)_{\hat{g}}+k,m,\nu^{l_{y-1},l_y}_{\hat{g}}(k))\cdot q(\nu^{l_{y-1},l_y}_{\hat{g}}(k),m)+\\ \nonumber
& \left.-\sum_{j=i}^{u}\sum_{k=1}^{\psi(l_{j-1},l_{j})_{\hat{g}}}n(\pi_d+\sum_{y=i}^{j-1}\psi(l_{y-1},l_{y})_{\hat{g}}+k,m,\nu^{l_{y-1},l_{y}}_{\hat{g}}(k))\cdot q(\nu^{l_{y-1},l_{y}}_{\hat{g}}(k),m)\right) \nonumber
\end{align}
\begin{align}\label{eq:phi2_ins}
\phi_2(l_{\hat{i}_{\overline{l}}-1},\overline{l},l_{\hat{i}_{\overline{l}}}) = & \max_{g \in G_{l^*,l_{\hat{i}_{\overline{l}}}}} \left( 1-\sum_{m \in \mathcal{M}}\alpha_{md}\right)\cdot \psi(l^*,l_{\hat{i}_{\overline{l}}})- \left( \sum_{m \in \mathcal{M}}\alpha_{md} \sum_{k=1}^{\psi(l^*,l_{\hat{i}_{\overline{l}}})_g}n(\pi_d+k,m,\nu^{l^*,l_{\hat{i}_{\overline{l}}}}_g(k))q(\nu^{l^*,l_{\hat{i}_{\overline{l}}}}_g(k),m) \right)+\\
& -\phi_1(l_{\hat{i}_{\overline{l}}-1},\overline{l},l_{\hat{i}_{\overline{l}}}) \nonumber
\end{align}
\vspace{-1cm}
\end{table*}

As reported in Algorithm \ref{alg:greedy}, the proposed greedy algorithm considers as inputs the sets of UAVs, time epochs and locations associated to deliveries and to the depot, as well as the UAVs' battery capacity and transmission capacity on the onboard radio system. Moreover, for every delivery, it takes as input the associated payload weight and time window, whereas for every UAV it takes as input the residual payload capacity available for parcels $\Delta(d) \leq Y$ and the parameters $\alpha_{md}$ (defined below). Note that the residual payload capacity of a UAV depends on the onboard equipment (e.g., antenna and monitoring system). In our greedy approach, the equipment of each UAV is chosen a priori\footnote{In section \ref{sec:results}, it will be shown that loading every UAV with the necessary equipment to perform all the additional missions in $\mathcal{M}$ leads to close-to-the-optimum performance.}. Therefore, the factors $\alpha_{md}$ are defined as follows: if UAV $d$ does not carry the equipment required to accomplish task $m$, then $\alpha_{md}=0$, otherwise $\alpha_{md}=\alpha_m$ , where $\alpha_m$ are predefined system weights such that $\sum_{m \in \mathcal{M}} \alpha_m \leq 1$. Finally, the time and energy required to travel between every location pair, the mission service requirements per location and epoch, the mission work quality per location, and the amount of data per epoch generated by each mission are provided as inputs (line 1).

Deliveries are ordered (line 2) depending on their urgency, i.e., the expiration of the delivery time window $b(p)$. Then UAVs are considered one at a time\footnote{In the case of a fleet of UAVs characterized by heterogeneous equipment, the criterion adopted for UAV selection can be designed according to the characteristics of the specific scenario. E.g., if delivery locations and their surroundings generate service requests for additional missions as well, then UAVs carrying the equipment required to fulfil such additional missions should be picked first. Conversely, if the areas with highest service requirements for additional missions do not include delivery points, then UAVs carrying no additional equipment should be selected first, since their residual payload capacity ($\Delta$) is higher and thus they can carry a larger amount of items to be delivered.} and the associated lists $TourL_d$ and $TourA_d$, respectively indicating the locations and the arcs included in their tours, are initialized. The depot $l^*$ is set as tour departure location. Moreover, their current battery consumption $e_d$, carried load $h_d$, and flight time $\pi_d$ are initialized to $0$ (line 4). 

Then, for every UAV, the ordered list of (yet unserved) deliveries is scanned one element at a time, the best route $\overline{g}$ to reach the delivery location is identified and feasible deliveries are assigned to the UAV (lines 5-11). For the selection of the best route $\overline{g}$, to incorporate the fulfillment of additional missions, we adopt the multi-objective approach described in \citep{zografos2004heuristic}: the arc to be traversed to reach the delivery location is chosen by minimizing a weighted function of traveling time and mission accomplishment levels associated to the arc, as per Eq. \eqref{eq:g_ins}, where $l_1=\overline{l}$ and $l_2=\hat{l}$ and parameters $\alpha_{md}$ are used as weights: the closer $\alpha_{md}$ approaches $1$, the more predominant the accomplishment of mission $m$ becomes w.r.t. the minimization of the total time duration of the tour.
Note that a delivery is considered to be feasible (line 7) if:
\begin{itemize}
\item[{\em (i)}] its delivery location $\hat{l}$ can be reached before the time window expiration and the UAV can return to the depot before the end of the considered time horizon, i.e., if it holds that $\pi_d+\psi(\overline{l},\hat{l})_{\overline{g}}\leq b(p)$ and that $\pi_d+\psi(\overline{l},\hat{l})_{\overline{g}} +\psi(\hat{l},l^*)_{g^*}\leq|\mathcal{K}|$, where $g^*\in G_{\hat{l},l^*}$ indicates the shortest route from location $\hat{l}$ to the depot;
\item[{\em (ii)}]  the residual battery capacity is sufficient to allow the flight to the delivery location and the return to the depot, i.e., if  $e_d+e(\overline{l},\hat{l})_{\overline{g}}+e(\hat{l},l^*)_{g^*} \leq E$;
  \item[{\em (iii)}] the residual payload capacity is sufficient to accommodate the item to be delivered, i.e., if $h_d+w(p)\leq \Delta(d)$. 
\end{itemize}
 
After the assignment of a delivery, the traversed arc and the delivery location are added to the UAV's tour. Moreover, traveling time, consumed energy, and residual payload capacity are updated (line 8) and the delivery is added to the set of served ones. Finally, the mission service requirements per epoch of every location $n(k,m,l)$ traversed to reach location $\hat{l}$ starting from location $\overline{l}$ are recomputed by means of the \texttt{update} routine (line 9) to reflect the residual needs after being served by the current UAV. The update criteria depend on the maximum transmission capacity of the onboard wireless interface of the UAV, $T$, on the data transmission requirements of each mission, $s(m)$, and on the parameters $\alpha_{md}$. More in detail, if the total data rate required in location $l$ at time $k$ to fully accomplish all the additional missions does not exceed $T$, then $n(k,m,l)$ are set either to 0, if the mission quality parameters are sufficiently high, or decreased by the amount $q(k,m)$ for every mission $m \in \Mcal$ (lines 4-6). Otherwise, the maximum transmission capacity is partitioned among the missions, proportionally to the product $\alpha_{md}s(m)$ (i.e., the data rate required by mission $m$ weighted by the importance of fulfilling mission $m$). %Note that missions with $\alpha_{md}>0$ are served first (lines 8-10), whereas the residual transmission capacity is dedicated to missions with $\alpha_{md}=0$, proportionally to their required data rates (lines 11-13). 
During the routine executions, the values of variables $\mu(d,m,k)$ are also computed according to the mission fulfillment share of every UAV.

When the delivery list scan is concluded, the best arc to return to the depot is selected according to Eq. \eqref{eq:g_ins}, where $l_1=\overline{l}$ and $l_2=l^*$, providing that the conditions $ e_d+e(\overline{l},l^*)_{g} \leq E$ and $\pi_d +\psi(\hat{l},l^*)_{g}\leq|\mathcal{K}|$ are satisfied; parameters $n(k,m,l)$ are again updated and the UAV tour is terminated (lines 12-17).

If some unserved deliveries remain after considering every available UAV, then the algorithm stops and notifies that it was unable to find a feasible solution (lines 18-19). Otherwise, the routine \texttt{conn\_ check} is executed to perform connectivity checks among UAVs. If a UAV cannot directly communicate with the cellular network, i.e., $t(L(d,k),\Omega)=0$, nor with another UAV having cellular network connectivity, i.e., $ \not\exists \ d' \in \Dc \colon$ $ d' \neq d $ $\wedge $ $t(L(d,k),L(d',k)>0)$ $\wedge t(L(d',k),\Omega)>0$, then no data generated by any mission can be transmitted; therefore, variable $\mu(d,m,k)$ is set to 0 for every mission $m \in \Mcal$.
Finally, the tour assigned to every UAV is returned (lines 20-21). 

On the contrary, if some UAVs remain unused, they  can perform additional missions, even if they are not assigned to any delivery mission. 
Thus, the algorithm can be executed again after enlarging the set of payload items $\mathcal{P}$ by generating fake delivery payloads  
(i.e., with $w(p)=0$) and no restriction on the delivery time window (i.e., $a(p)=1$ and $b(p)=|\mathcal{K}|$). 
The delivery locations $L(p)$ can be chosen randomly or placed where $n(k,m,L(p))$ have the highest values.

\subsection{Insertion-based heuristic}

\begin{algorithm}[tb]
 \caption{Insertion-based Algorithm}
    \label{alg:insertion}
\begin{algorithmic}[1]
\begin{small}
\STATE on input of $\mathcal{D}$; $\mathcal{K}$; $\mathcal{M}$; $\mathcal{L}$; $\hat{\mathcal{P}}$; $E$; $T$; $w(p),a(p),b(p), f(p) \forall p \in \hat{\mathcal{P}}$; $l^*$; $\psi(l,l')_g,e(l,l')_g,\nu^{ll'}_g, \forall l=f(p),l'=f'(p') \in \Lc, g \in G_{ll'}$; $\Delta(d),$ $\alpha_{md},$ $ n(k,m,l),$ $ q(l,m)$, $s(m)$ $ \forall m \in \mathcal{M},$ $ d \in \mathcal{D},$ $ l \in \mathcal{L}$; 
\FORALL {$d \in \mathcal{D}$}
	\STATE $TourL_d \leftarrow [l^*]$, $TourA_d \leftarrow []$, $e_d \leftarrow 0, \pi_d \leftarrow 0, h_d \leftarrow 0$
\ENDFOR
\STATE  $d \leftarrow 1$
\WHILE{$d \leq |\mathcal{D}| \wedge \hat{\mathcal{P}} \neq \oslash $}
\IF{$TourA_d$ is empty}
	\STATE $\overline{p} \leftarrow \argmin b(p)$, $\overline{l} \leftarrow l(\overline{p})$
	\STATE find best route $\overline{g}$ from $l^*$ to $\overline{l}$ 
	\STATE Append location $\overline{l}$ to $TourL_d$ and arc $(l^*,\overline{l})_{\overline{g}}$ to $TourA_d$
	\STATE $e_d \leftarrow e_d+e(\overline{l},l^*)_{\overline{g}}, \pi_d \leftarrow \pi_d+\psi(\overline{l},l^*)_{\overline{g}}, h_d \leftarrow h_d+w(p), \hat{\mathcal{P}} \leftarrow \hat{\mathcal{P}} \setminus {p}$
	\STATE find best route $\overline{g}$ from location $\overline{l}$ to location $l^*$ 
	\STATE Append location $l^*$ to $TourL_d$ and arc $(\overline{l},l^*)_{\overline{g}}$ to $TourA_d$
\ELSE
 	\FORALL {$l(p) \in \mathcal{L} \colon p \in \hat{\mathcal{P}}$}
 	\STATE $u \leftarrow $\texttt{len}$(TourL_d)$
 		\FORALL {$i \in \{1,...,u$\}}
 			\STATE $(l_{i-1},l_{i})_{\hat{g}} \leftarrow TourA_d[i]$
 			\STATE compute $\phi_1(l_{i-1},\overline{l},l_{i})$ 
 		\ENDFOR
 		\STATE $\hat{i}_{\overline{l}} \leftarrow \argmin_{i \in \{1,..,u\}} \phi_1(l_{i-1},\overline{l},l_{i})$
 		\STATE compute $\phi_2(l_{\hat{i}_{\overline{l}}-1},\overline{l},l_{\hat{i}_{\overline{l}}})$
 	\ENDFOR
 	\IF{$\max_{\overline{l}} \phi_2(l_{\hat{i}_{\overline{l}}-1},\overline{l},l_{\hat{i}_{\overline{l}}}) \geq 0$}
 		\STATE $\hat{l} \leftarrow \argmax_{\overline{l}} \phi_2(l_{\hat{i}_{\overline{l}}-1},\overline{l},l_{\hat{i}_{\overline{l}}})$
 		\STATE Insert location $\overline{l}$ in $TourL_d$ and arc $(l^*,\overline{l})_{\overline{g}}$ in $TourA_d$ after position $i$
		\STATE $e_d \leftarrow e_d+e(\overline{l},l^*)_{\overline{g}}, \pi_d \leftarrow \pi_d+\psi(\overline{l},l^*)_{\overline{g}}, h_d \leftarrow h_d+w(p), \hat{\mathcal{P}} \leftarrow \hat{\mathcal{P}} \setminus {p}$
	\ELSE
		\STATE $d \leftarrow d+1$,$\pi_d \leftarrow 0$
		\FORALL{$(\overline{l},\hat{l})_{\overline{g}} \in TourA_d$}
	\STATE $\forall m \in \Mcal, k \in \Kc \ \mu(d,m,k)  \leftarrow $\texttt{update}($\mathcal{K}$, $\mathcal{M}$, $\mathcal{L}$, $T$, $\pi_d$, $\psi(\overline{l},\hat{l})_{\overline{g}}$, $\nu_{\overline{g}}^{\overline{l}\hat{l}}$, $\alpha_{md}$, $ n(k,m,l)$, $ q(l,m)$, $s(m)$ $ \forall m \in \mathcal{M}$, $l \in \mathcal{L}$, $k \in \Kc$); $\pi_d \leftarrow \pi_d+\psi(\overline{l},\hat{l})_{\overline{g}}$
	\ENDFOR
 	\ENDIF
\ENDIF
\ENDWHILE
\IF{$\hat{\mathcal{P}} \neq \oslash$ }
 \RETURN no feasible solution found
\ELSE
\STATE $\forall d \in \Dc, m \in \Mcal, k \in \Kc \ \mu(d,m,k) \leftarrow$ \texttt{conn\_ check}($\mathcal{K}$; $\mathcal{D}$; $\mathcal{M}$; $\mathcal{L}$; $TourA_d \forall d \in \Dc$; $t(l,l') \forall l,l' \in \mathcal{L}$; $\mu(d,m,k) \forall d \in \Dc, m \in \Mcal, k \in \Kc$)
\RETURN {$\mu(d,m,k),TourL_d,TourA_d  \forall d \in \mathcal{D}, m \in \Mcal, k \in \Kc$}
\ENDIF
\end{small}
\end{algorithmic}
\end{algorithm}

Similarly to the greedy heuristic, the insertion-based heuristic aims at sequentially building the tours of each UAV by adding one delivery location at a time. However, here insertions are allowed at any point of the tour, whereas in the greedy heuristic new deliveries can be appended only at the end of a UAV's current tour.

As reported in Algorithm \ref{alg:insertion}, the insertion-based heuristic takes as input the same data of the greedy algorithm (line 1). The algorithm considers the UAVs sequentially and it first initializes the lists of tour locations and arcs, and the variables indicating the current battery consumption $e_d$, carried load $h_d$, and flight time $\pi_d$ (line 3).

If no delivery has yet been scheduled, as tour initialization criterion the insertion of the delivery with earliest deadline $\overline{p}$, located in $\overline{l}=L(\overline{p})$, is used (lines 8-13 - this criterion was chosen among those proposed in \citep{solomon1987algorithms}).
The route $\overline{g}$ used to travel from the depot $l^*$ to location $\overline{l}$ is selected (lines 9-11) as per Eq.\,\eqref{eq:g_ins}, where $l_1=l^*$ and $l_2=\overline{l}$, provided that the following conditions are satisfied:
\begin{itemize}
\item[{\em (i)}]  the energy availability constraint, i.e., $ e_d+e(l^*,\overline{l})_{g}+e(\overline{l},l^*)_{g^*} \leq E $;
\item[{\em (ii)}] the delivery time window satisfaction, i.e., $\pi_d+$ $\psi(l^*,\overline{l})_{g}$ $\leq b(p)$;
\item[{\em (iii)}] the condition on the return to depot, i.e., $\pi_d+\psi(l^*,\overline{l})_{g}$ $+\psi(\overline{l},l^*)_{g^*}$ $\leq|\mathcal{K}|$ where $(\overline{l},l^*)_{g^*}$ is the shortest route from $\overline{l}$ to the depot;
   \item[{\em (iv)}] the constraint on the maximum payload weight, i.e., $ h_d+w(p)\leq \Delta(d)$.
\end{itemize}
Next, the backward route to return from location $\overline{l}$ to the depot is selected with analogous procedure (lines 12-13), though in this case no constraints on the payload maximum weight nor on delivery time windows satisfaction need to be imposed to ensure the feasibility of the selected route. 

Then the algorithm iteratively operates as follows. Let \texttt{len} be an operator that, on input of a vector, returns the number of its elements. We define $u=$\texttt{len}($TourL_d$), where $TourL_d=[l_0,l_1,...,l_u]$ is the current route, with $l_0,l_u=l^*$ (line 16).
For each unserved delivery, the best insertion position in the UAV's tour $\hat{i}_{\overline{l}} \in \{1,..,u\}$ is evaluated by minimizing the function $\phi_1(l_{i-1},\overline{l},l_{i})$ as per Eq. \eqref{eq:phi1_ins}, where $\hat{g}$ is the route from $l_{i-1}$ to $l_{i}$ currently included in $TourA_d$ (lines 17-20).
Again, the insertion of the detour is considered as infeasible if either:
\begin{itemize}
\item[{\em (i)}]  inserting the detour required to reach location $\overline{l}$ in the tour leads the overall energy consumption to exceed the UAV battery capacity;
\item[{\em (ii)}]  the arrival epoch of the UAV at each delivery location does not meet the delivery time window constraint of the corresponding delivery task;
\item[{\em (iii)}] the time necessary to return to the depot exceeds $|\Kc|$.
 \end{itemize}
Note also that the computation of $\phi_1(l_{i-1},\overline{l},l_{i})$ takes into account the variations in the mission satisfaction levels per each epoch of the tour after the insertion of the candidate location $\overline{l}$, as service requirements may vary from epoch to epoch and inserting a new delivery point in the tour postpones the completion of the successive deliveries already inserted in the tour.  

Once the value $\hat{i}_{\overline{l}}=\argmin_{i \in 1,..,m} \phi_1(l_{i-1},\overline{l},l_{i})$ has been found (line 21), to choose the best unserved delivery to be inserted in the tour, the function $\phi_2(l_{\hat{i}_{\overline{l}}-1},\overline{l},l_{\hat{i}_{\overline{l}}})$ is computed as per Eq. \eqref{eq:phi2_ins} for every unserved delivery (line 22). Such function quantifies the improvements in the mission satisfaction levels obtained by adding delivery $\overline{l}$ in the current tour, as opposed to direct service of delivery $\overline{l}$ in a new, dedicated tour starting from the depot. If $\max_{\overline{l}} \phi_2(l_{\hat{i}_{\overline{l}}-1},\overline{l},l_{\hat{i}_{\overline{l}}}) \geq 0$ (line 24), it means that inserting the delivery in the current tour improves the mission satisfaction levels more than serving the delivery by means of a new tour. Therefore, delivery $\hat{l}=\argmax_{\overline{l} \in \Lc_u} \phi_2(l_{\hat{i}_{\overline{l}}-1},\overline{l},l_{\hat{i}_{\overline{l}}})$ is added to the current tour (lines 25-27) and the insertion procedure is repeated from the start. Otherwise, a new tour is initialized (line 28-29) and parameters $n(k,m,l)$ are recomputed via the \texttt{update} routine based on the locations visited during the previous tour.
The algorithm ends when all the delivery tasks are inserted in a tour, or when no more UAVs are available.
In this case, if there are remaining unserved deliveries, the algorithm notifies that no feasible solution can be found (lines 36-41).

\subsection{Complexity analysis}

If we consider only the parcel delivery tasks and no additional missions, the problem described in \Sec{model} can be modeled as a Vehicle Routing Problem
with Time Windows (VRPTW), which has been extensively studied in the
literature (a survey on heuristic and meta-heuristic approaches to VRPTW can be found in \citep{cordeau2000vrp}). However, since the existing studies do not consider multitasking vehicles such as our UAVs, the complexity of the surveyed heuristic approaches cannot be directly compared to that of our proposed algorithms.

The complexity of Algorithm \ref{alg:greedy} is $O(|\mathcal{D}| \cdot |\hat{\mathcal{P}} | \cdot \xi \cdot |\mathcal{M}| \cdot |\mathcal{K}|)$, as it is dominated by the two nested for loops at lines 3 and 5 and the minimization at line 6 within the inner loop.
Conversely, the complexity of Algorithm \ref{alg:insertion} is $O(|\mathcal{D}|\cdot |\hat{\mathcal{P}}^2 | \cdot \xi^2 \cdot |\mathcal{M}|^2\cdot |\mathcal{K}|^2)$. It follows that the complexity of both algorithms depends linearly on the number of UAVs, but Algorithm \ref{alg:greedy} (resp. Algorithm \ref{alg:insertion}) exhibits linear (respectively quadratic) dependency on the remaining input sizes.

\subsection{Analysis of worst-case ratio}\label{sec:bounds}
%\section{Bounds on Coverage and Monitoring (CM)}

%As it will be shown in the numerical assessment section, the proposed heuristic approaches yield results close to the optimal ones (and, thus, effective) for small scale scenarios. 
%As the calculation of the optimal solutions is computationally infeasible for large scale scenarios (due to the time requirements for solving the ILP model), 
In this section, we derive performance bounds under some simplifying assumptions. It should be clarified that we are interested in bounding $\Theta$, i.e., the satisfaction of the service demands generated by additional missions (defined as per Eq.\,\eqref{eq:obj}) that can be achieved by any solution that successfully supports all delivery missions. Consequently, in deriving the bounds it will be assumed that the delivery tasks are met (at some time epoch) and the focus will be on the derivation of upper and lower bounds of $\Theta$, respectively indicated as $\overline{\Theta}$ and $\underline{\Theta}$. In what follows, we focus on problem instances that admit at least one feasible solution.
Moreover, we do not assume any specific delivery time windows, i.e., $a(p)=1$ and $b(p)=|\mathcal{K}|$ $\forall p \in \hat{\mathcal{P}}$. This way, maximum flexibility in choosing the delivery epochs is allowed, so that no limitation on the satisfaction of the additional missions is imposed, e.g. by forcing some delivery locations to be visited earlier than others.

Let $I(k,m,l)$ be a binary parameter defined as follows:
\begin{equation}
\begin{footnotesize}
I(k,m,l)=\left\{
                \begin{array}{ll}
                  1 \ \mathtt{if} \ n(k,m,l)>0 \ \wedge \\ \quad \quad (k < \psi_{\overline{g}}(l^*,l) \vee k > |\Kc| -\psi_{\overline{g}}(l,l^*))\\
                  
                  0 \ \mathtt{otherwise}
                \end{array}
                \right.            
\end{footnotesize}
\end{equation}
where $\psi_{\overline{g}}(l^*,l)$ is the duration of the shortest path from the depot to location $l$. The parameter indicates if mission $m$ in location $l$ requires any service at epoch $k$ and if such location can be feasibly reached by a UAV, assuming that the UAV starts its trip at the first epoch of the time horizon from the depot $l^*$, follows the shortest trajectory to $l$, and returns to the depot by the last epoch.
Moreover, let $\mathcal{X}(k)_{\text{ord}}=[n_1(k,m,l),$ $n_2(k,m,l),$ $\dots,$ $n_{|\Mcal|\cdot |\Lc|}(k,m,l)]$ denote the ordered list of $n(k,m,l)$ for a given epoch $k$, in ascending order, and let $J_k=j \in \{1,\cdots,|\Mcal|\cdot |\Lc|\} \colon$ $ \sum_{i=1}^j n_i(k,m,l)\cdot I_i(k,m,l)\leq |\Dc|\cdot T $ $\wedge \sum_{i=1}^{j+1} n_i(k,m,l)\cdot I_i(k,m,l)> |\Dc|\cdot T$. This way, $J_k$ counts the maximum number of missions that can be fully satisfied at every time epoch, considering the limitation imposed by the maximum transmission capacity $T$ that each UAV can provide. Note that, if $\sum_{i=1}^j n_i(k,m,l)\cdot I_i(k,m,l) <|\Dc|\cdot T $, the $(J_k+1)$-th element of $\mathcal{X}(k)_{\text{ord}}$ constitutes a partially satisfied mission. Therefore, the overall satisfaction $\Theta$ is ensured to be strictly lower than $J_k+1$.
$\overline{\Theta}$ can then be expressed as:
\begin{small}
\begin{equation}\label{eq:upperC}
\overline{\Theta} =   \sum_{k \in \Kc}\min \left(J_k+1, \sum_{m \in \Mcal, l \in \Lc}I(k,m,l)\right)
\end{equation}
\end{small}
\begin{figure*}
\centering
\includegraphics[width=.8\textwidth]{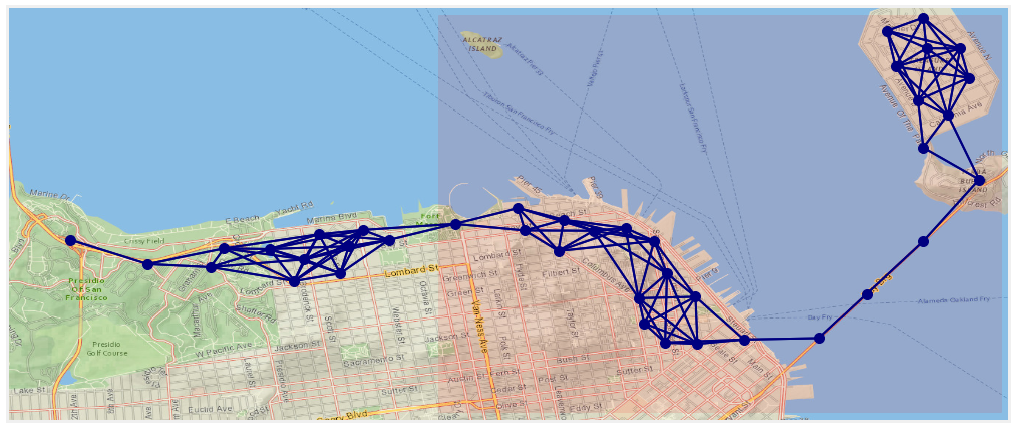}
\caption{
The reference topology we consider. Blue dots correspond to locations in~$\Lc$. Blue lines connect locations between which UAVs can travel in one epoch. The shadowed area corresponds to the small-scale topology we use for our comparison against the optimum.
    \label{fig:topo}
%    \vspace{-.5cm}
} %caption
\end{figure*}

%\begin{equation}
%\label{eq:upperC}
%\textcolor{red}{cm^u = |\Kc_e| ~ \{ min (\sum_{z\in\Zc} c_{max} , |\Dc | c^d ) + min (\sum_{z\in\Zc} m_{max}, |\Dc | m^d ) \}}
%\end{equation}

Under the aforementioned conditions, $\underline{\Theta}$ may be obtained by considering the briefest cycle $S\cal$ (where the cycle duration is measured in time epochs) that passes through the depot and all delivery locations. 
Note that, by definition, the briefest cycle cannot include hovering epochs. 
If the duration of $S\cal$, $|S\cal|$, exceeds $|\Kc|$, more that one UAV is needed  to reach all delivery locations; we denote by $\rho_e$  the minimum number of required UAVs, due to the UAVs' energy constraints. 
Furthermore, if the briefest cycle $S\cal$ cannot satisfy the specific time intervals restraining the time of deliveries, more than one UAV is needed to reach all delivery locations in a timely fashion. Let $\rho_t$ be the minimum number of required UAVs, meeting those time constraints. 
Finally, let $R=\max (\rho_e, \rho_t)  \geq 1$be the minimum number of UAVs required to satisfy all delivery missions. Based on triangular inequality arguments, it is evident that
$|S\cal| \leq $ $\sum_{i=1}^{R} |S_i\cal|$, where $|S_i\cal|$ denotes the duration of the path followed by the $i^{th}$ UAV. 

Moreover, let $[N^{m}_{\text{min}}, N^{m}_{\text{max}}]$ be the range of values of $n(k,m,l)$ for a given mission $m$; we define $\mathcal{N}_{\text{ord}}=[N^{m_1}_{\text{max}},$ $N^{m_2}_{\text{max}},$ $\dots, N^{m_{\Mcal}}_{\text{max}}]$ as the ordered list of $N^{m}_{\text{max}}$, in descending order. Then $\underline{\Theta}$ can be calculated as:
\begin{footnotesize}
\begin{equation}
\label{eq:lower2}
\underline{\Theta} = | S | \cdot \min \left(\sum_{i =1}^{|\Mcal|}\min \left( 1, \frac{\max(T-\sum_{j=1}^{i-1} N^{m_{j}}_{\text{max}},0) }{N^{m_i}_{\text{max}}}\right),|\Mcal|\right)
\end{equation}
\end{footnotesize} 
which quantifies the minimum satisfaction level that a single UAV can ensure to every mission while visiting any location at any time epoch. More in detail, if the capacity $T$ of one UAV is sufficient to satisfy the requirements per epoch of all the additional missions in any location during any epoch, then the (minimum) number of fully-served additional missions per epoch is $|\mathcal{M}|$. Otherwise, the term $\min \left( 1, \frac{\max(T-\sum_{j=1}^{i-1} N^{m_{j}}_{\text{max}},0) }{N^{m_i}_{\text{max}}}\right)$ counts the minimum satisfaction level that can be achieved under the worst-case assumption. That is, considering that in the location traversed by the UAV at epochs $1,2,...,|S|$ all the additional missions exhibit their highest service requirement $N^{m_{i}}_{\text{max}}$ and that capacity $T$ is partitioned among them so that the most capacity-demanding missions are served first, until exhaustion of $T$.

Notice that $| S\cal |$ is lower-bounded by twice the distance (in epochs) between the depot location $l^*$ and the farthest delivery location, $l_f$; that is, 
$| S | \geq \min_{g \in G_{l^*l_f}}2\psi_g ( l^*,l_f )$. Consequently, if $| S |$ is unknown, a looser lower bound could be obtained by replacing $| S\cal |$ with $\min_{g \in G_{l^*l_f}}2\psi_g ( l^*,l_f )$ in \eqref{eq:lower2}. 

%An alternative lower bound could be obtained by noting that at least $\frac{| S\cal |}{2}$ distinct locations must be traversed while traveling along the briefest cycle. Let $\mathcal{N}^{m}_{\text{ord}}$ denote the ordered list of $n(k,m,l)$ for each mission $m$, in ascending order. Then an alternative lower bound can be obtained by the following.

%\begin{equation}
%\label{eq:lower3}
%\underline{\Theta}' = 2 \sum_{i=1}^{ {| S\cal |} \over {2 }}\sum_{m \in M} \mathcal{N}^{m}_{\text{ord}}(i) 
%\end{equation} 

Conversely, for the cases in which $R >1$ distinct cycles are needed to meet all deliveries in a timely manner (where each cycle is covered by one of the $R$ UAVs), an alternative lower bound could be derived as follows. Let $\{l_1, l_2, ... , l_{R-1} \}$ denote the locations of the closest $R-1$ delivery locations to the depot. Then, the minimum cumulative flying time of the UAVs is given by: 
$$\sum_{i=1}^{R-1}2 \min_{g \in G_{ l^* l_i}}\psi ( l^* , l_i )+\min_{g \in G_{ l_d l_f}} \psi ( l_d , l_f)$$
since the farthest location $l_f$ must be reached by at least one UAV and the remaining $R-1$ UAVs will necessarily serve at least one delivery during their trip. Thus, the duration of their trip must exceed the time required to fly from the depot to the $R-1$ nearest delivery locations and return. In this case, $| S\cal | $ can be replaced in \Eq{lower2} with the latter quantity.

Finally, an upper bound on the worst-case performance ratio can be obtained by computing the ratio of Eq.\,\eqref{eq:upperC} to Eq.\,\eqref{eq:lower2}. %(or Eq. \eqref{eq:lower3}).
%Note that, if using Eq. \eqref{eq:lower2}, such bound is finite as long as either $N^{m}_{\text{min}}$ is strictly positive for al least one mission $m$. Conversely, if \eqref{eq:lower3} is used, then the ratio is finite as long as at most $\frac{|S|}{2}-1$ locations have $n(k,m,l)=0$ for some $m$ and some $k$.

\section{Reference Scenarios}\label{sec:scenario}

\begin{table}
\caption{Simulation parameters
\label{tab:params}
}
\begin{tabularx}{1\columnwidth}{|p{4cm}|p{1.75cm}|p{1.8cm}|}
\hline
Quantity & Small  $\quad$ scenarios & Large $ \quad$ scenarios \\ \hline
Number of locations, $|\Lc|$ & 28 & 40 \\ \hline
Number of deliveries, $|\hat{\Pc}|$& 7 & 20 \\ \hline
Number of UAVs, $|\Dc|$& 10--15 & 20--30 \\ \hline
Number of epochs, $|\Kc|$& \multicolumn{2}{l|}{20} \\ \hline
Epoch duration [min] & \multicolumn{2}{l|}{10} \\ \hline
UAV max. speed [km/h]& \multicolumn{2}{l|}{6} \\ \hline
UAV battery capacity [Wh], $E$ & \multicolumn{2}{l|}{230} \\ \hline
\end{tabularx}
\end{table}

For our reference scenarios,
we consider a flood in the area of the city of San Francisco, as depicted in \Fig{topo}. The flooding event has been simulated through the software Hazus~\citep{hazus}, and, given the disaster area, we identify~$|\Lc|=40$ locations. UAVs have to perform a total of~$|\hat{\Pc}|=20$ deliveries of blood or medicine packs, due at randomly-selected locations out of the $\Lc$ set (the $f$-parameters), over a time window of 10~epochs for medicine packs and 5~epochs for blood packs (our~$a$- and $b$-parameters).
Locations associated with a delivery have zero elevation, all others have an elevation of 50~meters. Also notice how~\citep{noi-workshop19} elevation changes have a limited impact on the overall power consumption.
We generate a total of 20~scenarios, sharing the same topology and disaster, but with different delivery time windows.

\begin{figure*}
\centering
\subfigure[\label{fig:optonly-perf}]{
    \includegraphics[width=.47\textwidth]{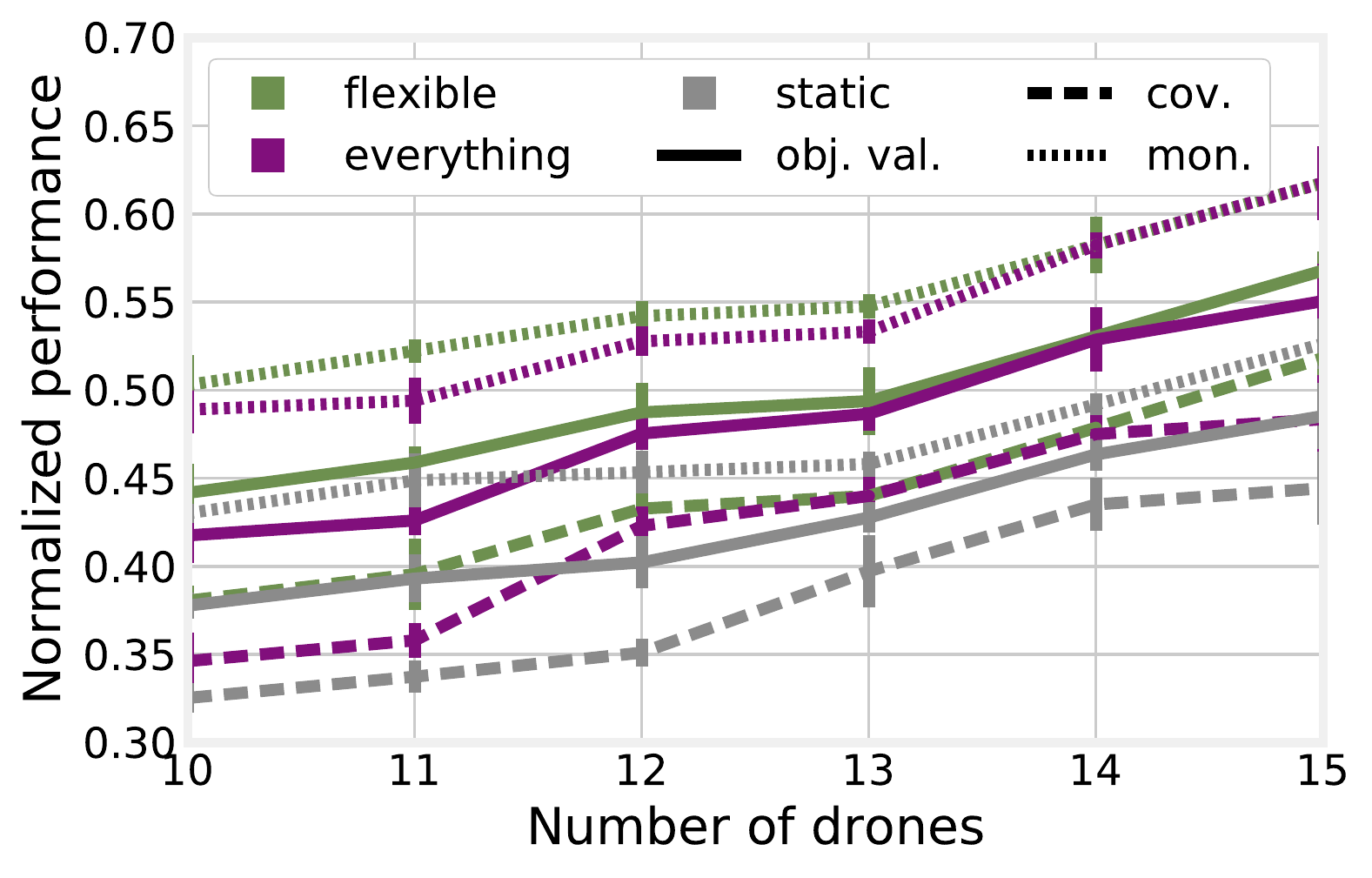}
} %subfigure
\subfigure[\label{fig:optonly-weight}]{
    \includegraphics[width=.47\textwidth]{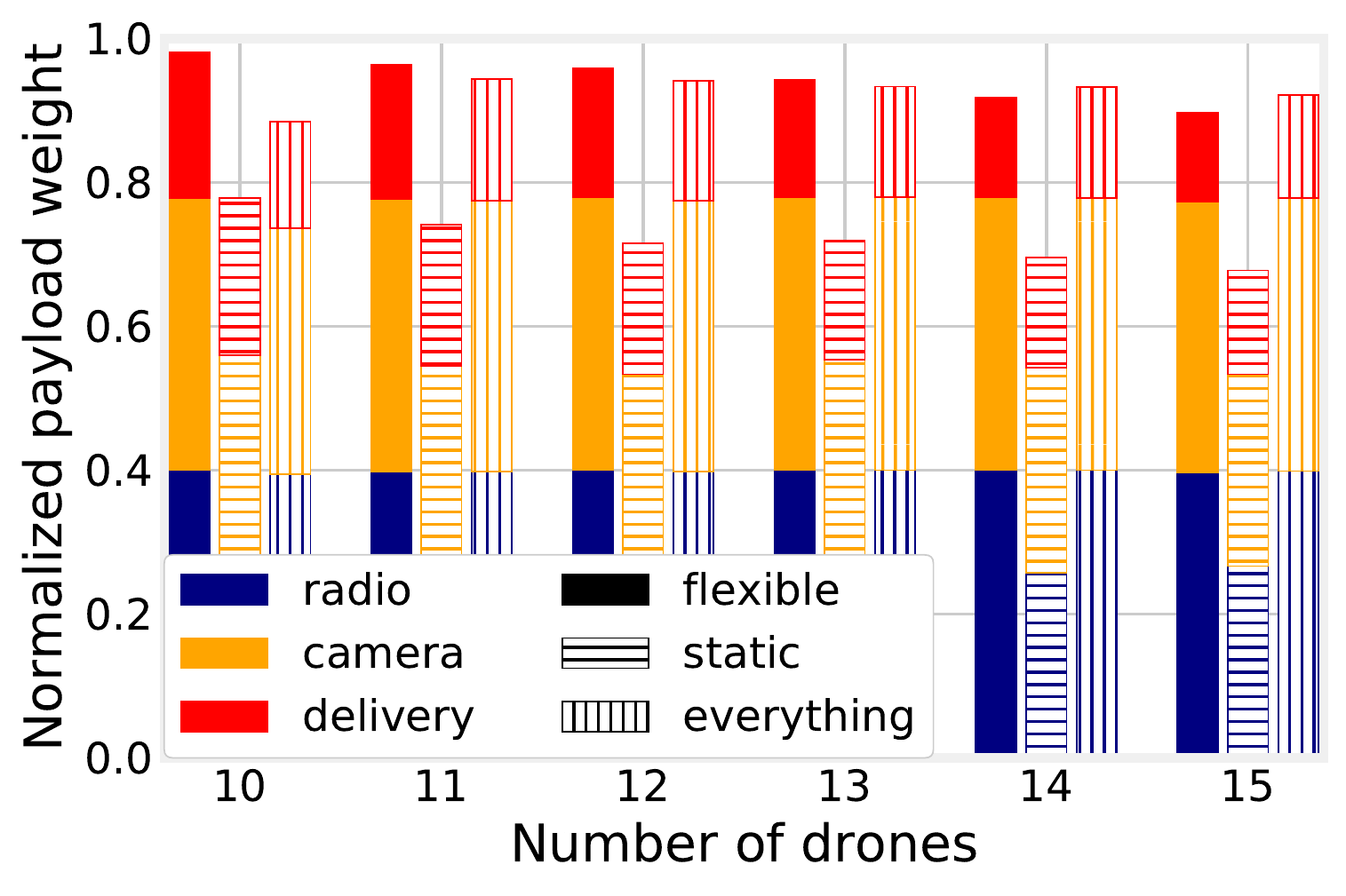}
} %subfigure
\caption{
    Small-scale scenarios, optimal decisions: performance (a) and payload (b) when payload assignment is flexible, static, and when all UAVs always carry radio and camera (``everything''). Performance is normalized by the total demand, payload by the total capacity~$Y$, and used energy by the battery capacity~$E$.
    \label{fig:optonly}
} %caption
\end{figure*}
\begin{figure*}
\centering
\subfigure[\label{fig:optheur-perf}]{
    \includegraphics[width=.47\textwidth]{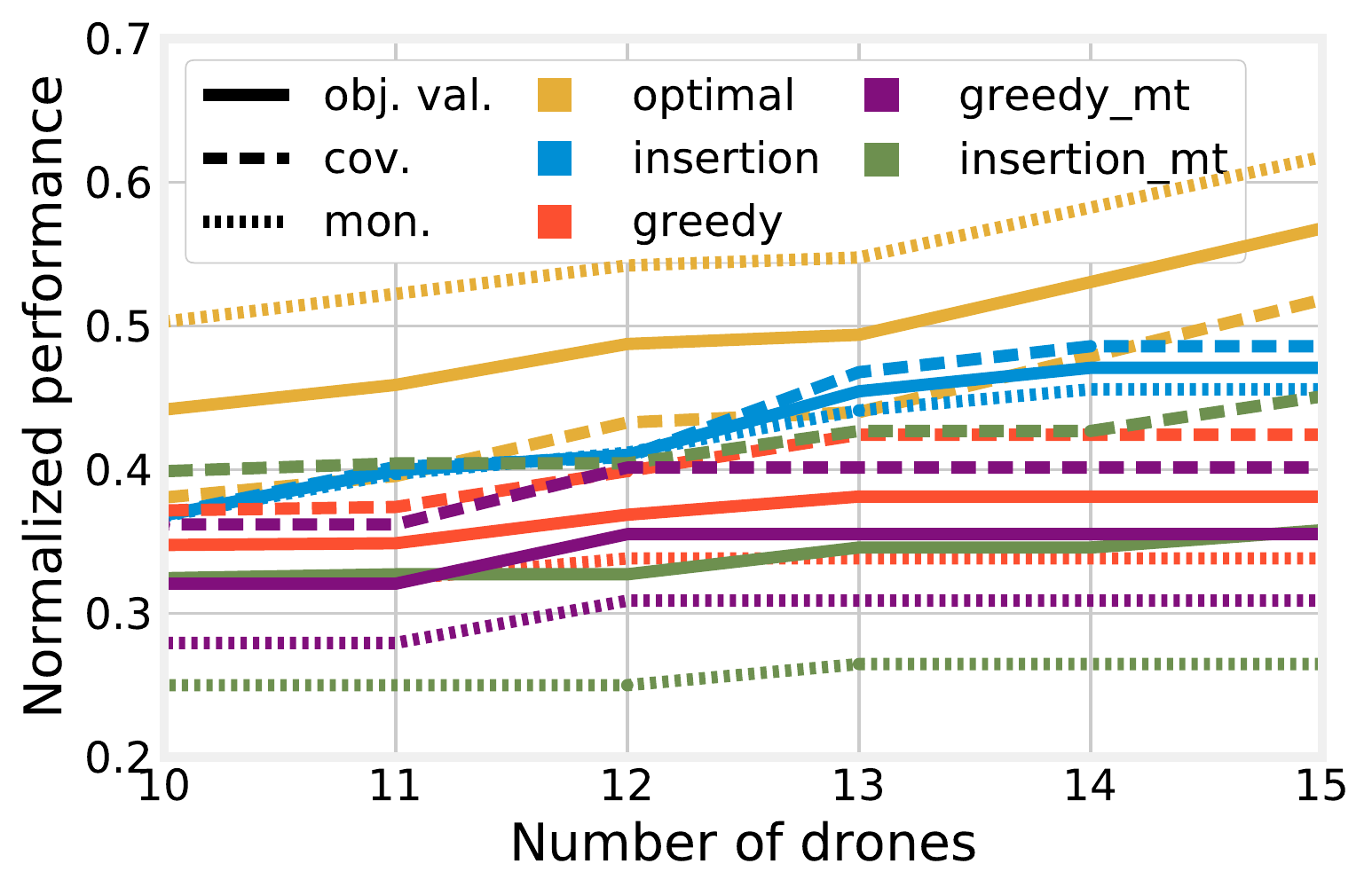}
} %subfigure
\subfigure[\label{fig:optheur-scatter}]{
    \includegraphics[width=.47\textwidth]{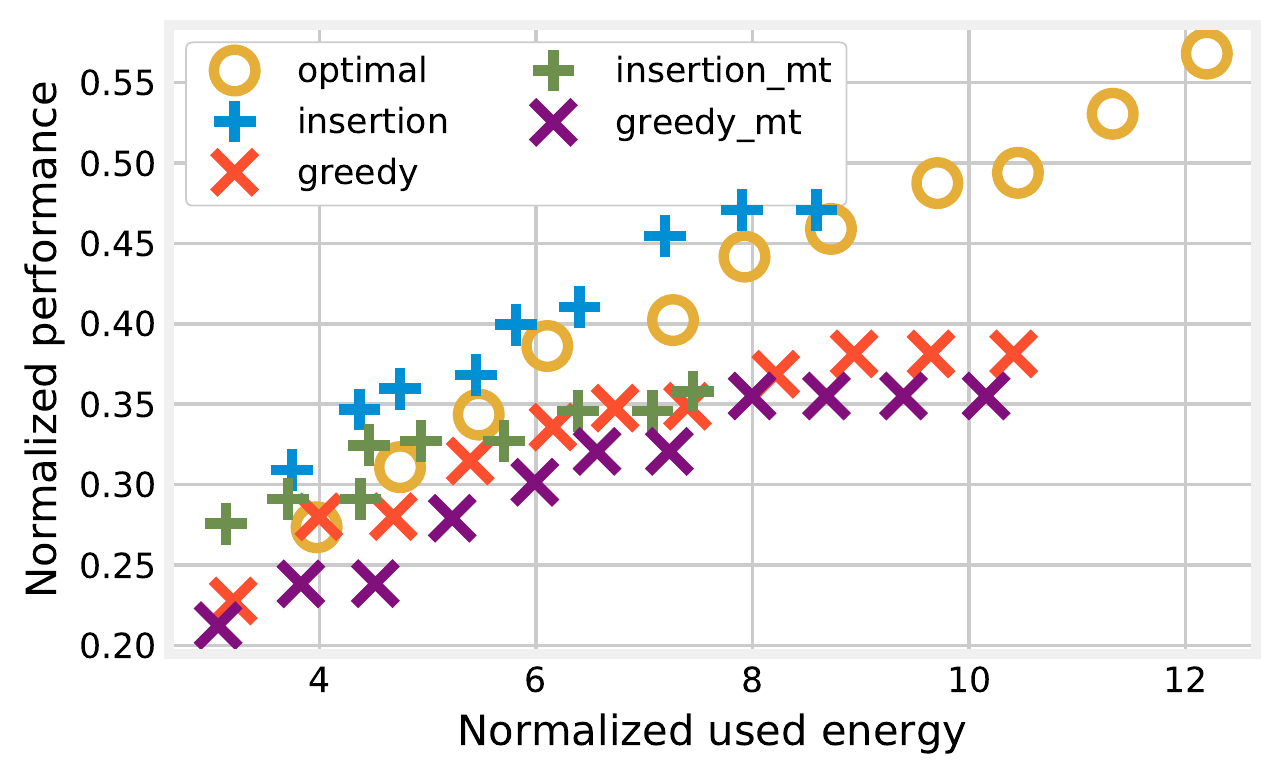}
} %subfigure
\caption{
    Small-scale scenarios: performance (a) and energy/performance trade-off (b) yielded when decisions are made optimally (yellow lines/markers), through the insertion heuristics (blue), and through the greedy heuristics (red). Performance is normalized by the total demand, payload by the total capacity~$Y$, and used energy by the battery capacity~$E$.
    Confidence intervals, omitted from the figure for sake of readability, are narrower than 0.08 in all cases.
    \label{fig:optheur}
} %caption
\end{figure*}
\begin{figure*}
\centering
\subfigure[\label{fig:twoheur-perf}]{
    \includegraphics[width=.47\textwidth]{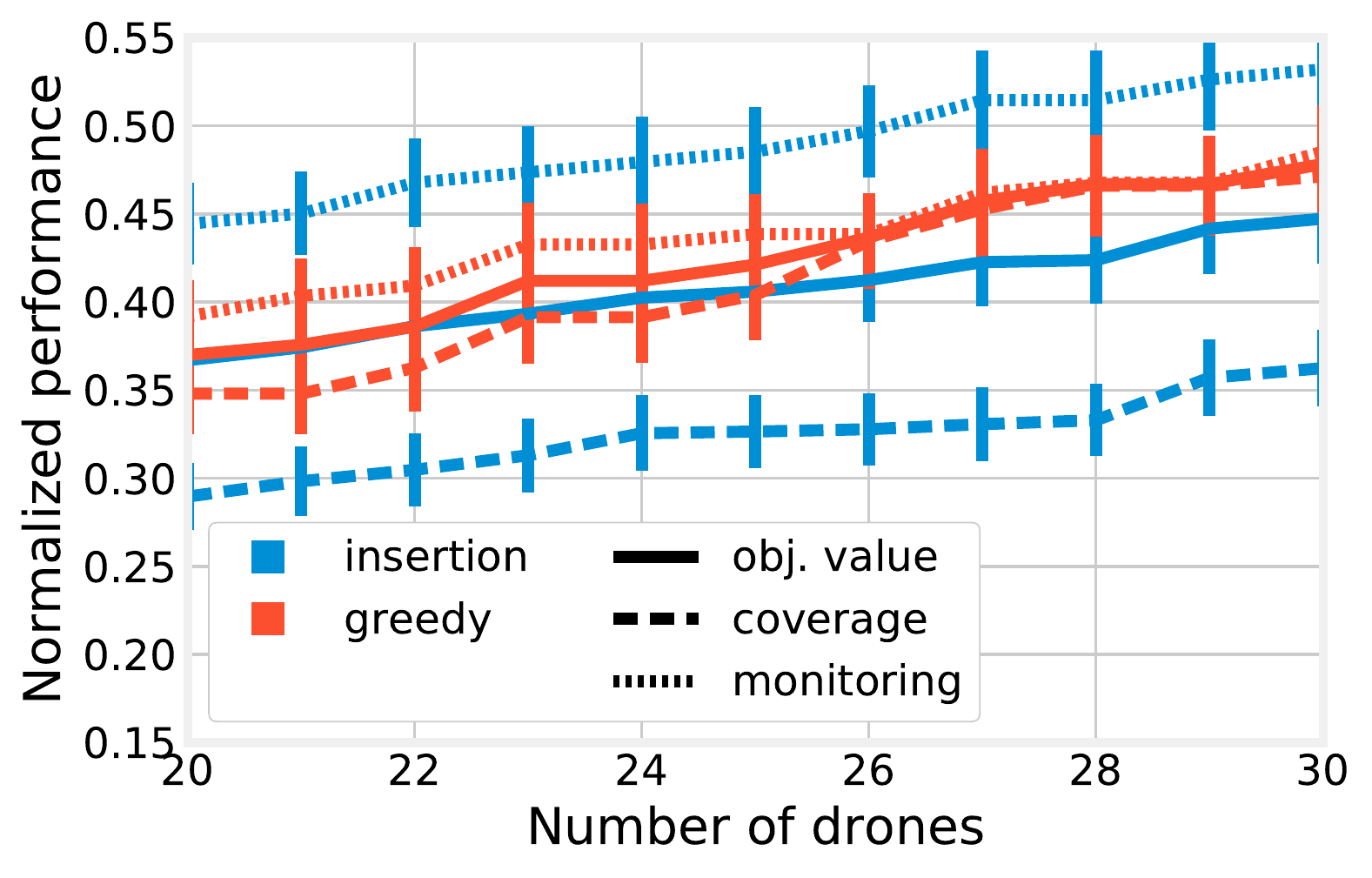}
} %subfigure
\hspace{-0.2cm}
\subfigure[\label{fig:twoheur-scatter}]{
    \includegraphics[width=.47\textwidth]{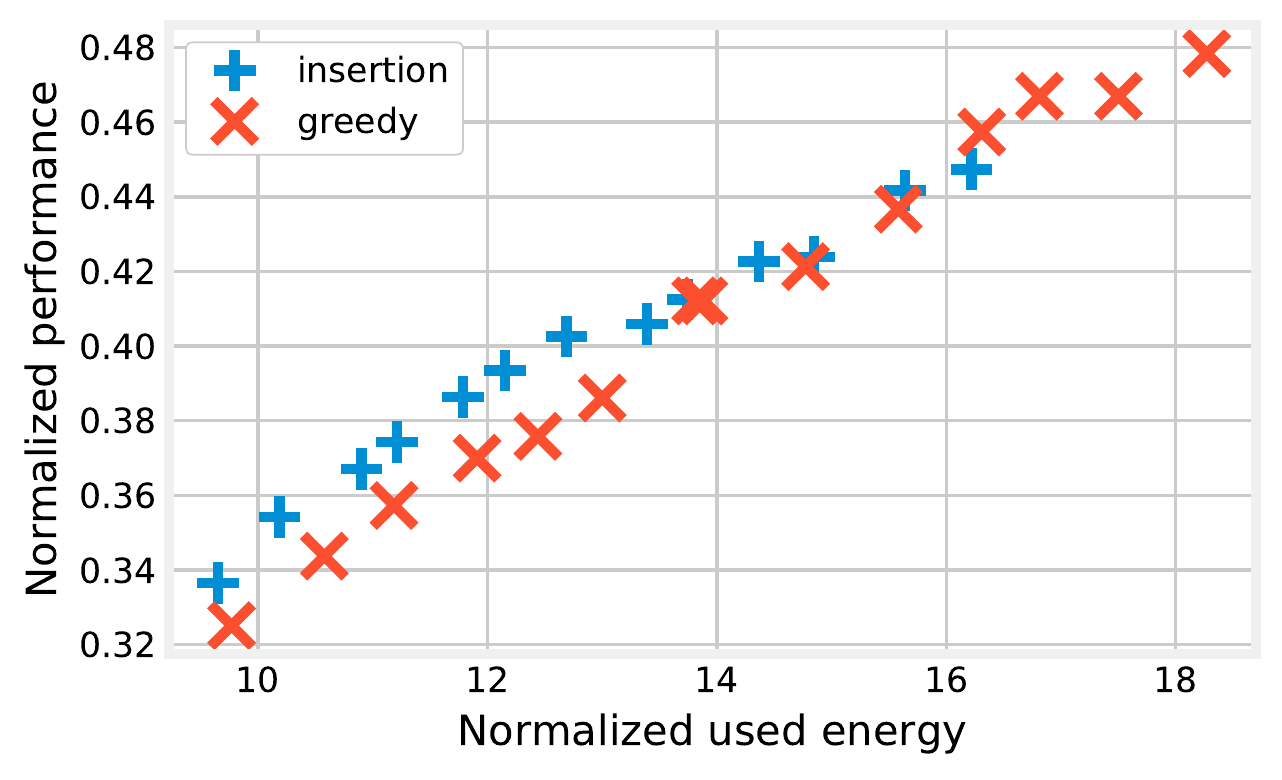}
} %subfigure
\caption{
    Large-scale scenarios: performance (a) and energy/performance trade-off (b) for the insertion-based and greedy heuristics. Performance is normalized by the total demand, payload by the total capacity~$Y$, and used energy by the battery capacity~$E$.
    \label{fig:large}
} %caption
\end{figure*}

%\begin{figure}
%\centering
%    \includegraphics[width=.47\textwidth]{img_dic19/ext_parstudy_performance_timevar.png}
%\caption{
%    Large-scale scenario: effect of the parameters on the performance of the insertion-based heuristics. Performance is normalized by the total demand, payload by the total capacity~$Y$, and used energy by the battery capacity~$E$.
%    \label{fig:parstudy-perf}
%} %caption
%\end{figure}

UAVs can also perform $|\Mcal|=2$ additional missions: $i)$ providing network coverage for users escaping from the disaster, whose mobility is simulated through the MatSim simulator~\citep{matsim}, as detailed in~\citep{noi-workshop19}; $ii)$ video monitoring, e.g., to assess the level of the flooding and damage in a certain area.

The quantity of needed service (the $n$-parameters) is determined as follows. For the coverage mission, we use the values computed in~\citep{noi-workshop19}, based on the expected flow of vehicles. 
For video monitoring, a subset of 50\% randomly-selected locations are deemed to need service, i.e., $n(l)=1$, while all others have~$n(l)=0$. Coverage and monitoring missions require additional payloads, respectively, the  radio~\citep{ettus} and the camera system~\citep{kurz2011real}, each weighting 1~kg. The maximum throughput values achievable between any two locations, i.e., the $t$-parameters, are obtained with reference to LTE micro-cells through the methodology in~\citep{noi-workshop19}.

We consider a set of UAVs of variable cardinality, whose features mimic those of lightweight Amazon UAVs~\citep{xu2017design}. Specifically, they have an empty weight of~$W=4~\text{kg}$, and a maximum payload of~$Y=2.5~\text{kg}$. They are equipped with a battery of capacity~$E=230~\text{Wh}$, and the energy consumed to fly between locations is~$e(l_1,l_2)=3.125~\text{Wh/km/kg}$. As a result, the range of a UAV carrying its maximum payload is around~$\frac{E}{e(C+W)}=11.3~\text{km}$. Interestingly, such a figure matches the 10-km range envisioned for lightweight UAVs in~\citep[Tab.~1]{stolaroff2018energy}.
Finally,  we consider~$|\Kc|=20$ epochs, each corresponding to 10~minutes.

The main scenario features and simulation parameters are summarized in \Tab{params}.

\section{Numerical Assessment}\label{sec:results}

We first investigate the relationship between payload and equipment assignment to UAVs and the resulting coverage and monitoring performance. To this end, in \Fig{optonly} we focus on the small-scale scenarios, derive the optimal solution via the off-the-shelf commercial solver Gurobi, and compare the following three alternatives:
\begin{itemize}
    \item {\bf flexible}: no constraint about the $\omega$-variables is imposed. Hence, UAVs carry whatever payloads maximize the objective \Eq{obj};
    \item {\bf static}: one third of UAVs always carry a radio, one third a camera, and one third neither of them;
    \item {\bf everything}: all UAVs always carry both a camera and a radio.
\end{itemize}
%Beyond the constraints on the payload to carry (if any), all other decisions are made optimally.

\Fig{optonly-perf} shows the performance obtained by the above payload assignment strategies, as the number of UAVs changes. 
The flexible assignment is always associated with very good performance for both coverage (dashed lines) and monitoring (dotted lines), as well as the values of the objective function in \Eq{obj} (solid lines). More interestingly, when all UAVs carry both cameras and radios (``everything''), the performance is almost the same as in the flexible case, within~7\% for any number of UAVs. Conversely, the static payload assignment is associated with a noticeably worse performance, up to 22\%~off.

It can be noticed that, even when the number of UAVs is significant, the normalized performance remains below 60\% in most cases. This is due to the fact that, in all cases, coverage and monitoring have lower priority compared to deliveries, and are performed (so to say) on a best-effort basis. This also explains the frequently different performance associated with coverage and monitoring, in spite of being weighted equally in the objective function: UAVs will perform whichever task happens to be closer to their destination, catching, in a way, the low-hanging fruit.

\Fig{optonly-weight} representing how the UAVs' payload capacity is used, sheds further light on the decisions made under the three strategies. 
We can observe that, when the payload assignment is flexible (solid bars), almost all UAVs carry both a camera and a radio, just like in the ``everything'' case (vertical hatches). Under the static assignment (horizontal hatches), there is often a significant portion (20--30\%) of the UAVs' payload capacity that is left unused, except when the number of UAVs is very small. 
This contradicts the intuition that carrying a radio and/or a camera is somehow detrimental to the UAVs' ability to perform delivery missions, and motivates us to set our heuristics so that all UAVs always carry both a camera and a radio.

Still focusing on  the small-scale scenarios, in \Fig{optheur} we compare the optimal performance to that of the two heuristics introduced in \Sec{heuristic}. The values of the parameters $\alpha_{md}$ leading to highest $\Theta$ have been identified beforehand by performing a sensitivity analysis with step $0.1$, whereas $\xi$ is set to 10. Moreover, we report results for the case $\alpha_{md}=0 \ \forall m \in \mathcal{M}, d \in \mathcal{D}$: under this setting, the computations in Eq. \eqref{eq:g_ins}-\eqref{eq:phi2_ins} only consider the travel duration between locations, disregarding the amount of service provided for additional missions. These results can therefore be considered as a benchmark case where only the timely accomplishment of delivery task is taken into account.
\Fig{optheur-perf} shows how the performance of both heuristics compares to the optimal one. 
Heuristics with optimized choice of the $\alpha_{md}$ parameters tend to perform similarly (or, indeed, better) in terms of coverage, and worse in terms of monitoring. 
Among the two heuristics, the insertion-based one always outperforms, by 10--20\%, the greedy heuristic  in both coverage and monitoring. Setting parameters $\alpha_{md}$ to 0 leads to significant performance degradations, and in this case the two heuristic approaches exhibit similar performance.

\Fig{optheur-scatter} summarizes the energy/performance trade-offs achieved by each alternative. Each marker therein corresponds 
to a value of the number of UAVs: its x- and y-position represent, respectively, the total used energy 
and the average performance, computed summing the normalized performance for coverage and monitoring, and then dividing 
the result by two. It is possible to observe how, for the same number of UAVs, the optimal solution tends 
to consume more energy (see, as an example, the rightmost yellow markers). 
Since UAVs weight virtually the same, such a difference is due to the fact that, under the optimal solution, 
UAVs make longer trips. Indeed, recall that under both heuristics UAVs only move from a delivery location to another along one of the $\xi$ shortest paths, which limits the length of their trips.

In \Fig{large}, we move to the large-scale scenarios, and compare the performance and behavior of the two heuristics. 
\Fig{twoheur-perf} shows that the relationship between their performance is more complex than in \Fig{optheur-perf}. Specifically, 
the insertion-based heuristic performs up to 20\%~better than the greedy one for coverage, 
but around 10\%~worse for monitoring. The reason becomes evident in \Fig{twoheur-scatter}, highlighting how the greedy 
heuristic yields longer trips (and it consumes more energy). Notice that, owing to the scenario size, the optimal solution could be computed.

%The issue of balancing the monitoring and coverage performance is tackled, as discussed in \Sec{heuristic}, through the parameters of the insertion-based heuristics. Changing them, it is possible to seek for a balanced performance, or to privilege one of the two tasks. The impact of parameter settings on the performance is shown in \Fig{parstudy-perf}, which highlights how different settings yield different trade-offs between coverage and monitoring. The figure further suggests how, by setting the appropriate parameters, the insertion-based heuristics is able to tackle different scenarios with different conditions.

It is important to stress that in the small scenarios making optimal decisions takes several hours on the Xeon-based server 
we used for our simulations, while the timings of the greedy and insertion-based heuristics ranged within $[0.49,1.46]$ seconds and $[19.70,137.99]$ seconds, respectively. In the large scenarios, timings ranged within $[3.03,7.98]$ seconds and $[218.29,795.77]$ seconds, respectively. As a reference, optimization times ranged within $[2820,32482]$~seconds.

Thus, we can conclude that the heuristics represent an excellent trade-off between computational complexity and decision quality.

\section{Conclusions}
We addressed the challenging problem of jointly planning the missions
of multitask UAVs and applied it to a post-disaster scenario. In
such cases, tasks are expected to be associated with a common
geographical area (i.e., the disaster area). Hence UAVs carrying out
such tasks would largely geographically overlap. %We show that assigning multiple tasks to UAVs can lead to savings in the number of UAVs required to carry out all the tasks, provided that the problem of jointly planning the multiple tasks iseffectively addressed. 
To this end, we developed an optimization
formulation and two heuristic approaches that effectively cope with
the computational complexity posed by the task. 

Our performance evaluation, carried out leveraging a
realistic model of a flood in the San
Francisco area, as well as realistic
parameters for the operational 
equipment and tasks,
confirmed our main intuition, that is, that the flexibility of multi-task UAVs allows better performance than single-purpose ones, in spite of the higher weight they have to carry. Furthermore, our heuristics  have shown remarkably good performance, consistently close to the optimum in spite of their very low complexity.

\bibliographystyle{IEEEtran}
\bibliography{sigproc}

\begin{thebibliography}{10}
\providecommand{\url}[1]{#1}
\csname url@samestyle\endcsname
\providecommand{\newblock}{\relax}
\providecommand{\bibinfo}[2]{#2}
\providecommand{\BIBentrySTDinterwordspacing}{\spaceskip=0pt\relax}
\providecommand{\BIBentryALTinterwordstretchfactor}{4}
\providecommand{\BIBentryALTinterwordspacing}{\spaceskip=\fontdimen2\font plus
\BIBentryALTinterwordstretchfactor\fontdimen3\font minus
  \fontdimen4\font\relax}
\providecommand{\BIBforeignlanguage}[2]{{%
\expandafter\ifx\csname l@#1\endcsname\relax
\typeout{** WARNING: IEEEtran.bst: No hyphenation pattern has been}%
\typeout{** loaded for the language `#1'. Using the pattern for}%
\typeout{** the default language instead.}%
\else
\language=\csname l@#1\endcsname
\fi
#2}}
\providecommand{\BIBdecl}{\relax}
\BIBdecl

\bibitem{erdelj2017help}
M.~Erdelj, E.~Natalizio, K.~R. Chowdhury, and I.~F. Akyildiz, ``Help from the
  sky: Leveraging {UAV}s for disaster management,'' \emph{IEEE Pervasive
  Computing}, vol.~16, no.~1, p. 24–32, Jan. 2017.

\bibitem{saeed2017realistic}
A.~Saeed, A.~Abdelkader, M.~Khan, A.~Neishaboori, K.~A. Harras, and A.~Mohamed,
  ``On realistic target coverage by autonomous drones,''
  \emph{arXiv:1702.03456}, 2017.

\bibitem{bamburry2015drones}
D.~Bamburry, ``Drones: Designed for product delivery,'' \emph{Design Management
  Review}, vol.~26, no.~1, pp. 40--48, 2015.

\bibitem{fotouhi2018survey}
A.~Fotouhi, H.~Qiang, M.~Ding, M.~Hassan, L.~G. Giordano, A.~Garcia-Rodriguez,
  and J.~Yuan, ``Survey on {UAV} cellular communications: Practical aspects,
  standardization advancements, regulation, and security challenges,''
  \emph{IEEE Communications Surveys \& Tutorials}, vol.~21, no.~4, pp.
  3417--3442, 2019.

\bibitem{ackerman2018medical}
E.~Ackerman and E.~Strickland, ``Medical delivery drones take flight in east
  {A}frica,'' \emph{IEEE Spectrum}, vol.~55, no.~1, pp. 34--35, 2018.

\bibitem{lee2017optimization}
J.~Lee, ``Optimization of a modular drone delivery system,'' in \emph{2017
  Annual IEEE International Systems Conference (SysCon)}.\hskip 1em plus 0.5em
  minus 0.4em\relax IEEE, 2017, pp. 1--8.

\bibitem{kurz2011real}
F.~Kurz, D.~Rosenbaum, J.~Leitloff, O.~Meynberg, and P.~Reinartz, ``Real time
  camera system for disaster and traffic monitoring,'' in \emph{2011
  International Conference on Sensors and Models inPhotogrammetry and Remote
  Sensing (SMPR)}, 2011.

\bibitem{otto2018optimization}
A.~Otto, N.~Agatz, J.~Campbell, B.~Golden, and E.~Pesch, ``Optimization
  approaches for civil applications of unmanned aerial vehicles ({UAV}s) or
  aerial drones: A survey,'' \emph{Networks}, vol.~72, no.~4, pp. 411--458,
  2018.

\bibitem{erdelj2016uav}
M.~Erdelj and E.~Natalizio, ``{UAV}-assisted disaster management: Applications
  and open issues,'' in \emph{2016 IEEE International conference on Computing,
  Networking and Communications (ICNC)}.\hskip 1em plus 0.5em minus 0.4em\relax
  IEEE, 2016, pp. 1--5.

\bibitem{mozaffari2018tutorial}
M.~Mozaffari \emph{et~al.}, ``{A tutorial on {UAV}s for wireless networks:
  Applications, challenges, and open problems},'' \emph{arXiv:1803.00680},
  2018.

\bibitem{mozaffari2015UAV}
M.~Mozaffari, W.~Saad, M.~Bennis, and M.~Debbah, ``Drone small cells in the
  clouds: Design, deployment and performance analysis,'' in \emph{2015 IEEE
  Global Communications Conference (GLOBECOM)}.\hskip 1em plus 0.5em minus
  0.4em\relax IEEE, 2015, pp. 1--6.

\bibitem{bor2016efficient}
R.~I. Bor-Yaliniz, A.~El-Keyi, and H.~Yanikomeroglu, ``Efficient 3-d placement
  of an aerial base station in next generation cellular networks,'' in
  \emph{2016 IEEE international conference on communications (ICC)}.\hskip 1em
  plus 0.5em minus 0.4em\relax IEEE, 2016, pp. 1--5.

\bibitem{reina2018multi}
D.~Reina, H.~Tawfik, and S.~Toral, ``Multi-subpopulation evolutionary
  algorithms for coverage deployment of {UAV}-networks,'' \emph{Ad Hoc
  Networks}, vol.~68, pp. 16--32, 2018.

\bibitem{zhao2018deployment}
H.~Zhao, H.~Wang, W.~Wu, and J.~Wei, ``Deployment algorithms for {UAV} airborne
  networks toward on-demand coverage,'' \emph{IEEE Journal on Selected Areas in
  Communications}, vol.~36, no.~9, pp. 2015--2031, 2018.

\bibitem{trotta2018joint}
A.~Trotta, M.~Di~Felice, F.~Montori, K.~R. Chowdhury, and L.~Bononi, ``Joint
  coverage, connectivity, and charging strategies for distributed {UAV}
  networks,'' \emph{IEEE Transactions on Robotics}, vol.~34, no.~4, pp.
  883--900, 2018.

\bibitem{bupe2015relief}
P.~Bupe, R.~Haddad, and F.~Rios-Gutierrez, ``Relief and emergency communication
  network based on an autonomous decentralized {UAV} clustering network,'' in
  \emph{2015 IEEE SoutheastCon}.\hskip 1em plus 0.5em minus 0.4em\relax IEEE,
  2015, pp. 1--8.

\bibitem{hironet}
L.~Ferranti, S.~D’Oro, L.~Bonati, E.~Demirors, F.~Cuomoy, and T.~Melodia,
  ``Hiro-net: Self-organized robotic mesh networking for internet sharing in
  disaster scenarios,'' in \emph{20th IEEE International Symposium on a World
  of Wireless, Mobile and Multimedia Networks, WoWMoM}.\hskip 1em plus 0.5em
  minus 0.4em\relax IEEE, 2019.

\bibitem{mkiramweni2018game}
M.~E. Mkiramweni, C.~Yang, J.~Li, and Z.~Han, ``Game-theoretic approaches for
  wireless communications with unmanned aerial vehicles,'' \emph{IEEE Wireless
  Communications}, vol.~25, no.~6, pp. 104--112, 2018.

\bibitem{mkiramweni2019survey}
M.~E. Mkiramweni, C.~Yang, J.~Li, and W.~Zhang, ``A survey of game theory in
  unmanned aerial vehicles communications,'' \emph{IEEE Communications Surveys
  \& Tutorials}, vol.~21, no.~4, pp. 3386--3416, 2019.

\bibitem{REINA201861}
D.~Reina, T.~Camp, A.~Munjal, and S.~Toral, ``Evolutionary deployment and local
  search-based movements of 0th responders in disaster scenarios,''
  \emph{Future Generation Computer Systems}, vol.~88, pp. 61 -- 78, 2018.

\bibitem{SANCHEZGARCIA2019129}
\BIBentryALTinterwordspacing
J.~Sánchez-García, D.~Reina, and S.~Toral, ``A distributed pso-based
  exploration algorithm for a uav network assisting a disaster scenario,''
  \emph{Future Generation Computer Systems}, vol.~90, pp. 129 -- 148, 2019.
  [Online]. Available:
  \url{http://www.sciencedirect.com/science/article/pii/S0167739X18303649}
\BIBentrySTDinterwordspacing

\bibitem{chmaj2015distributed}
G.~Chmaj and H.~Selvaraj, ``Distributed processing applications for
  {UAV}/drones: A survey,'' in \emph{Progress in Systems Engineering},
  H.~Selvaraj, D.~Zydek, and G.~Chmaj, Eds.\hskip 1em plus 0.5em minus
  0.4em\relax Cham: Springer International Publishing, 2015, pp. 449--454.

\bibitem{changchun2010research}
L.~Changchun, S.~Li, W.~Hai-bo, and L.~Tianjie, ``The research on unmanned
  aerial vehicle remote sensing and its applications,'' in \emph{2010 2nd
  International Conference on Advanced Computer Control}, vol.~2.\hskip 1em
  plus 0.5em minus 0.4em\relax IEEE, 2010, pp. 644--647.

\bibitem{garapati2017game}
K.~Garapati, J.~J. Rold{\'a}n, M.~Garz{\'o}n, J.~del Cerro, and A.~Barrientos,
  ``A game of drones: Game theoretic approaches for multi-robot task allocation
  in security missions,'' in \emph{Iberian Robotics conference}.\hskip 1em plus
  0.5em minus 0.4em\relax Springer, 2017, pp. 855--866.

\bibitem{alfeo2019enhancing}
A.~L. Alfeo, M.~G. Cimino, and G.~Vaglini, ``Enhancing biologically inspired
  swarm behavior: Metaheuristics to foster the optimization of uavs
  coordination in target search,'' \emph{Computers \& Operations Research},
  vol. 110, pp. 34--47, 2019.

\bibitem{zema2017unmanned}
N.~R. Zema, E.~Natalizio, and E.~Yanmaz, ``An unmanned aerial vehicle network
  for sport event filming with communication constraints,'' in \emph{BalkanCom
  2017 First International Balkan Conference on Communications and Networking},
  2017.

\bibitem{luo2015uav}
C.~Luo, J.~Nightingale, E.~Asemota, and C.~Grecos, ``A {UAV}-cloud system for
  disaster sensing applications,'' in \emph{2015 IEEE 81st Vehicular Technology
  Conference (VTC Spring)}.\hskip 1em plus 0.5em minus 0.4em\relax IEEE, 2015,
  pp. 1--5.

\bibitem{yoo2018drone}
W.~Yoo, E.~Yu, and J.~Jung, ``Drone delivery: Factors affecting the public’s
  attitude and intention to adopt,'' \emph{Telematics and Informatics},
  vol.~35, no.~6, pp. 1687--1700, 2018.

\bibitem{murray2015flying}
C.~C. Murray and A.~G. Chu, ``The flying sidekick traveling salesman problem:
  Optimization of drone-assisted parcel delivery,'' \emph{Transportation
  Research Part C: Emerging Technologies}, vol.~54, pp. 86--109, 2015.

\bibitem{7934790}
J.~{Lee}, ``Optimization of a modular drone delivery system,'' in \emph{2017
  Annual IEEE International Systems Conference (SysCon)}, April 2017, pp. 1--8.

\bibitem{peng2019hybrid}
K.~Peng, J.~Du, F.~Lu, Q.~Sun, Y.~Dong, P.~Zhou, and M.~Hu, ``A hybrid genetic
  algorithm on routing and scheduling for vehicle-assisted multi-drone parcel
  delivery,'' \emph{IEEE Access}, vol.~7, pp. 49\,191--49\,200, 2019.

\bibitem{krakowczyk2018developing}
D.~Krakowczyk, J.~Wolff, A.~Ciobanu, D.~J. Meyer, and C.-E. Hrabia,
  ``Developing a distributed drone delivery system with a hybrid behavior
  planning system,'' in \emph{2018 Joint German/Austrian Conference on
  Artificial Intelligence (K{\"u}nstliche Intelligenz)}.\hskip 1em plus 0.5em
  minus 0.4em\relax Springer, 2018, pp. 107--114.

\bibitem{ferrandez2016optimization}
S.~M. Ferrandez, T.~Harbison, T.~Weber, R.~Sturges, and R.~Rich, ``Optimization
  of a truck-drone in tandem delivery network using k-means and genetic
  algorithm,'' \emph{Journal of Industrial Engineering and Management (JIEM)},
  vol.~9, no.~2, pp. 374--388, 2016.

\bibitem{liu2018application}
J.~Liu, Z.~Guan, J.~Shang, and X.~Xie, ``Application of drone in solving last
  mile parcel delivery,'' \emph{Journal of Systems Science and Information},
  vol.~6, no.~4, pp. 302--319, 2018.

\bibitem{shiri2019massive}
H.~Shiri, J.~Park, and M.~Bennis, ``Massive autonomous uav path planning: A
  neural network based mean-field game theoretic approach,'' in \emph{2019 IEEE
  Global Communications Conference (GLOBECOM)}.\hskip 1em plus 0.5em minus
  0.4em\relax IEEE, 2019, pp. 1--6.

\bibitem{ghambari2018comparative}
S.~Ghambari, J.~Lepagnot, L.~Jourdan, and L.~Idoumghar, ``A comparative study
  of meta-heuristic algorithms for solving uav path planning,'' in \emph{2018
  IEEE Symposium Series on Computational Intelligence (SSCI)}.\hskip 1em plus
  0.5em minus 0.4em\relax IEEE, 2018, pp. 174--181.

\bibitem{scott2017drone}
J.~Scott and C.~Scott, ``Drone delivery models for healthcare,'' in \emph{2017
  50th Hawaii international conference on system sciences}, 2017.

\bibitem{malandrino_catania}
F.~Malandrino, C.~Rottondi, C.-F. Chiasserini, A.~Bianco, and I.~Stavrakakis,
  ``Multiservice {UAV}s for emergency tasks in post-disaster scenarios,'' in
  \emph{ACM MobiHoc workshop on innovative aerial communication solutions for
  FIrst REsponders network in emergency scenarios (iFIRE’19)}, 2019.

\bibitem{zeng2019energy}
Y.~Zeng, J.~Xu, and R.~Zhang, ``Energy minimization for wireless communication
  with rotary-wing {UAV},'' \emph{IEEE Transactions on Wireless
  Communications}, 2019.

\bibitem{man1996approximation}
M.~R. Garey and D.~S. Johnson, ``Approximation algorithms for bin packing
  problems: A survey,'' pp. 147--172, 1981.

\bibitem{solomon1987algorithms}
M.~M. Solomon, ``Algorithms for the vehicle routing and scheduling problems
  with time window constraints,'' \emph{Oper. research}, vol.~35, no.~2, pp.
  254--265, 1987.

\bibitem{zografos2004heuristic}
K.~G. Zografos and K.~N. Androutsopoulos, ``A heuristic algorithm for solving
  hazardous materials distribution problems,'' \emph{European Journal of
  Operational Research}, vol. 152, no.~2, pp. 507--519, 2004.

\bibitem{cordeau2000vrp}
J.-F. Cordeau and G.~d'{\'e}tudes et de recherche en analyse des~d{\'e}cisions
  (Montr{\'e}al~Qu{\'e}bec), \emph{The VRP with time windows}, 2000.

\bibitem{hazus}
{U.S. FEMA}. {Hazus program}. \url{https://www.fema.gov/hazus}.

\bibitem{noi-workshop19}
L.~{Chiaraviglio}, L.~{Amorosi}, F.~{Malandrino}, C.~F. {Chiasserini},
  P.~{Dell'Olmo}, and C.~{Casetti}, ``Optimal throughput management in
  uav-based networks during disasters,'' in \emph{IEEE INFOCOM 2019 - IEEE
  Conference on Computer Communications Workshops (INFOCOM WKSHPS)}, 2019, pp.
  307--312.

\bibitem{matsim}
A.~Horni, K.~Nagel, and K.~W. Axhausen, \emph{The multi-agent transport
  simulation MATSim}.\hskip 1em plus 0.5em minus 0.4em\relax Ubiquity Press
  London, 2016.

\bibitem{ettus}
{Ettus}. {N200 software radio datasheet}.
  \url{https://www.ettus.com/content/files/Ettus_N200-210_DS_Flyer_HR_2.pdf }.

\bibitem{xu2017design}
J.~Xu, \emph{Design perspectives on delivery drones}.\hskip 1em plus 0.5em
  minus 0.4em\relax RAND Corporation, 2017.

\bibitem{stolaroff2018energy}
J.~K. Stolaroff, C.~Samaras, E.~R. O’Neill, A.~Lubers, A.~S. Mitchell, and
  D.~Ceperley, ``Energy use and life cycle greenhouse gas emissions of drones
  for commercial package delivery,'' \emph{Nature communications}, vol.~9,
  no.~1, pp. 1--13, 2018.

\end{thebibliography}

\section*{Acknowledgment}
This work  was supported by the European Commission through the I-REACT project  (grant agreement n.~700256).

\end{document}